\documentclass[twocolumn,letterpaper,accepted=2026-06-05]{quantumarticle}
\pdfoutput=1

\usepackage{graphicx}

\usepackage[utf8]{inputenc}
\usepackage[english]{babel}
\usepackage[T1]{fontenc}
\usepackage{amsmath}
\usepackage{amssymb}
\usepackage{mathrsfs}

\usepackage{hyperref}
\usepackage{bm}
\usepackage{subfigure}
\usepackage[dvipsnames]{xcolor}

\newcommand{\beq}{\begin{equation}}
\newcommand{\eeq}{\end{equation}}
\newcommand{\beqnn}{\begin{equation*}}
\newcommand{\eeqnn}{\end{equation*}}
\newcommand{\bea}{\begin{eqnarray}}
\newcommand{\eea}{\end{eqnarray}}
\newcommand{\beann}{\begin{eqnarray*}}
\newcommand{\eeann}{\end{eqnarray*}}
\newcommand{\bes} {\begin{subequations}}
\newcommand{\ees} {\end{subequations}}

\newcommand{\ket}[1]{ | #1\rangle}

\newcommand{\ketbra}[2]{|#1\rangle\langle #2|}
\newcommand{\mB}{\mathcal{B}}

\newcommand{\mK}{\mathcal{K}}
\newcommand{\mL}{\mathcal{L}}

\newcommand{\mU}{\mathcal{U}}

\newcommand{\ident}{\openone}
\newcommand{\Tr}{\mathrm{Tr}}

\newcommand{\ignore}[1]{}

\usepackage{mdframed}

\usepackage{tabularray}

\begin{document}

\title{Temporal Coarse Graining for Classical Stochastic Noise in Quantum Systems}

\author{Tameem Albash}
\email{tnalbas@sandia.gov}
\affiliation{Center for Computing Research, Sandia National Laboratories, Albuquerque NM, 87185 USA}
\orcid{0000-0003-3916-3985}

\author{Steve Young}
\affiliation{Center for Computing Research, Sandia National Laboratories, Albuquerque NM, 87185 USA}

\author{N. Tobias Jacobson}
\affiliation{Center for Computing Research, Sandia National Laboratories, Albuquerque NM, 87185 USA}

\begin{abstract}
Simulations of quantum systems with Hamiltonian classical stochastic noise can be challenging when the noise exhibits temporal correlations over a multitude of time scales, such as for $1/f$ noise in solid-state quantum information processors. Here we present an approach for simulating Hamiltonian classical stochastic noise that performs temporal coarse-graining by effectively integrating out the high-frequency components of the noise. We focus on the case where the stochastic noise can be expressed as a sum of Ornstein-Uhlenbeck processes. Temporal coarse-graining is then achieved by conditioning the stochastic process on a coarse realization of the noise, expressing the conditioned stochastic process in terms of a sum of smooth, deterministic functions and bridge processes with boundaries fixed at zero, and performing the ensemble average over the bridge processes. For Ornstein-Uhlenbeck processes, the deterministic components capture all dependence on the coarse realization, and the stochastic bridge processes are not only independent but taken from the same distribution with correlators that can be expressed analytically, allowing the associated noise propagators to be precomputed once for all simulations. This combination of noise trajectories on a coarse time grid and ensemble averaging over bridge processes has practical advantages, such as a simple concatenation rule, that we highlight with numerical examples.
\end{abstract}
\maketitle
\section{Introduction}
The development of quantum information processors benefits from detailed modeling of underlying noise processes for error attribution of benchmark performance~\cite{Emerson2005,Emerson2007,Knill2008,Magesan2011,Mageson2012,Sheldon2016,Proctor2019,Nielsen2021,Proctor2025}, control pulse optimizations~\cite{Grace2012,Edmunds2020}, and error correction~\cite{Robertson2017,Tuckett2019,BonillaAtaides2021}. As error rates become smaller, it becomes important to develop simulation methods that are able to faithfully simulate the noisy dynamics over longer time scales in order to magnify the effects of noise. With primitive operations on the order 100's of nanoseconds, experiments of benchmarking protocols can be on the order of seconds~\cite{Andrews2019,Tanttu2024}, so wall-clock simulations of these experiments cover a broad range of timescales.

Accurate simulation of such systems can pose a serious challenge even in the case where the noise can be described as a classical stochastic process in the Hamiltonian \cite{Green2013,Crow2014,Rossi2017,Halataei2017,Cerfontaine2021,Hangleiter2021,Peng2022}. This is true, for example, for solid-state qubit systems~\cite{Loss1998,Petta2005b,Petta2005,Maune2012,Prance2012,Bukard2023} where the $T_1$ times for relaxation processes are sufficiently large compared to $T_2$ times such that they may be ignored~\cite{Tyryshkin2012,Kamyar2013}. These systems are challenging to simulate, even with this simplification, because they exhibit noise with temporally correlated noise over large time scales such as in the case of $1/f$ noise~\cite{Dutta1981,Paladino2014,Yoneda2018,RojasArias2024}. 
For example, brute-force integration of the Schr\"odinger equation necessitates a small time step to faithfully capture the high-frequency components of the noise and avoid aliasing, but this small step size can be orders of magnitude longer than the time needed to study the effect of slow drift~\cite{vanEnk2013,Fogarty2015,Klimov2018,Rudinger2019,Proctor2020,Tanttu2024} arising from the low frequency components. This broad range of relevant timescales results in high simulation cost.

\begin{figure*}[!t] %
   \centering
    \includegraphics[width=6in]{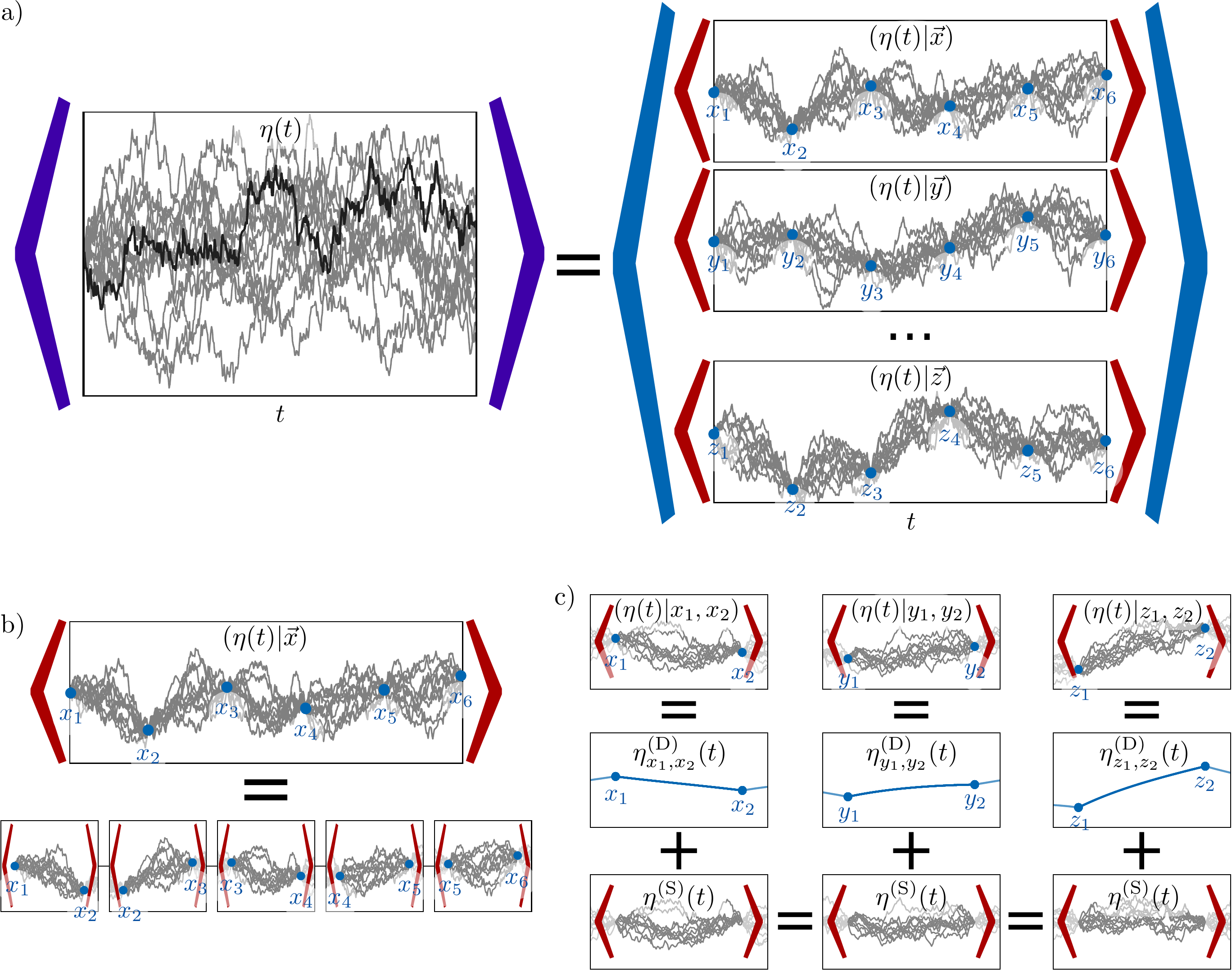} 
   \caption{Schematic of how the ensemble average is performed in our temporal coarse-graining approach. The vertical axis corresponds to the deviation of the stochastic process from some mean value, which we take to be zero. (a) The ensemble average over independent OU stochastic processes $\eta(t)$, denoted by the purple brackets, is implemented using two different ensemble averages. The first ensemble average (red bracket) is over OU processes conditioned on a coarse realization of the process (blue labeled points). For example, for a coarse realization given by $\vec{x} = (x_1, x_2, \dots)$ at the time points $t_1, t_2, \dots$ that need not be evenly spaced, the conditioned OU process is denoted as $(\eta(t) | \vec{x})$.  This is followed by an ensemble average over independent coarse noise realizations of the OU process (blue bracket). (b) The ensemble average over the conditioned OU processes satisfies a simple concatenation rule because each time segment between a pair of coarse time points is independent. (c) The conditioned OU process $(\eta(t)|x_1, x_2)$ can be expressed in terms of the sum of a deterministic function $\eta_{x_1,x_2}^{(\mathrm{D})}(t)$, depicted in blue, which depends on the coarse noise realization, and a zero-boundary bridge process $\eta^{(\mathrm{S})}(t)$, depicted in gray, that is independent of the coarse noise realization but depends on the temporal separation $t_2 - t_1$. The zero-boundary bridge processes for different coarse noise realizations are independent and identically distributed, and the ensemble average is performed over the zero-boundary bridge processes.}\label{fig:SummaryFigure}
\end{figure*}

In this work, we focus on the case of noise consisting of classical stochastic processes in the Hamiltonian $H(t)$ of the form:
\beq \label{eqt:H}
H(t) = H_{\mathrm{I}}(t) + H_{\mathrm{N}}(t) \ ,
\eeq
where $H_{\mathrm{I}}(t)$ denotes the ideal time-dependent Hamiltonian that generates the ideal dynamics and $H_{\mathrm{N}}(t)$ denotes the noise described by classical stochastic noise processes. In this context, one approach to circumvent the above multi-scale noise issue is to derive an effective quantum process description~\cite{Kraus1983,Nielsen2011} of the system dynamics over some relevant time scale using the standard Magnus~\cite{Magnus1954,Blanes2009} and cumulant expansion~\cite{Kubo1962,Kubo1963}. This is accomplished by performing an ensemble average over the stochastic processes under suitable assumptions about the noise. This approach can be recast in terms of filter functions~\cite{Kofman2001,Martinis2003,Uhrig2007,Cywinski2008,Clausen2010,Green2013}, as presented in Refs.~\cite{Cerfontaine2021,Hangleiter2021}, which provides a convenient formulation. The disadvantage of this approach though is that in the case of a sequence of unitary operations, i.e. $U(t_n,t_0) = \prod_{k=1}^n U_k(t_k,t_{k-1})$, such as those describing a quantum circuit, there is not a concatenation rule to calculate the quantum process except under the assumption that the ideal Hamiltonian $H_{\mathrm{I}}(t)$ is constant during each interval $[t_{k-1},t_k]$ or if certain second order Magnus expansion terms are ignored~\cite{Cerfontaine2021,Hangleiter2021}.

Here we present an alternative approach to tackling the multi-scale noise issue. We focus on stochastic processes that can be expressed as a sum of independent Ornstein-Uhlenbeck (OU) processes~\cite{Uhlenbeck1930}. While choosing to focus on OU processes, which are Markovian and Gaussian, may seem restrictive, sums of independent OU processes can give rise to non-Markovian Gaussian processes that capture physically relevant $1/f$ power spectral densities \cite{Kaulakys2005}. We nonetheless expect our approach to be applicable to other Markovian processes.

In our approach, temporal coarse-graining is achieved by conditioning each of our OU stochastic processes on a realization of the noise process on a coarse time grid. The conditioned stochastic process is then given by a sequence of  OU bridge processes~\cite{Goldys2008}, where each bridge process has its boundary values fixed by the coarse noise realization. Each OU bridge processes can be further expressed in terms of a deterministic function and an OU bridge process that is fixed at zero at its boundaries. Performing an ensemble average over the zero-boundary bridge process effectively integrates over the high frequency components of the noise. Furthermore, the independence of the zero-boundary bridge processes means the ensemble average dynamics corresponds to a composition of maps for each coarse time step, which gives a simple concatenation rule. The ensemble average over the noise is completed by taking an average over different realizations of the noise process on the coarse time grid. We illustrate this procedure in Fig.~\ref{fig:SummaryFigure}. 

This approach based on conditioning the stochastic process gives rise to a `hybrid' method involving a Monte Carlo aspect (generating noise realizations on the coarse time grid) and ensemble averaged dynamics. We note that our approach is also different from the approach of temporal coarse-graining proposed in Ref.~\cite{Gullans2024}, which relies on implementing process tomography sequentially in time. While our discussion focuses on Hamiltonian dynamics, the approach we describe can also be used for dynamics governed by a Gorini–Kossakowski–Sudarshan–Lindblad master equation~\cite{Gorini1976,Lindblad1976} with classical stochastic noise.

The manuscript is organized as follows. In Sec.~\ref{sec:Method} we give a detailed derivation of our approach using conditioned stochastic processes. In Sec.~\ref{sec:Results}, we present three examples to illustrate this approach.  In particular, we simulate repeated applications of a 3-qubit weight-2 parity check circuit encoded in 6 spins. This simulation involves a sequence of mid-circuit measurements, which requires temporal correlation to be tracked across the mid-circuit measurement, and is achieved without any additional overhead with our method. In Sec.~\ref{sec:efficiency}, we discuss the computational costs of our method in more detail. Finally, we conclude in Sec.~\ref{sec:conclusions}.
\section{Method} \label{sec:Method}
%
We proceed to give a detailed derivation of our approach for simulating the dynamics associated with a time-dependent Hamiltonian $H(t)$ of the form in Eq.~\eqref{eqt:H}. In Sec.~\ref{sec:Review}, we present a derivation that uses the standard ensemble average and follows Ref.~\cite{Hangleiter2021} closely. The initial part of this derivation, before the ensemble average is performed, is identical in our approach, and in Sec.~\ref{sec:BridgeEnsembleAverage} we present our approach to performing the ensemble average into two steps.
\subsection{Magnus and Cumulant Expansion Formalism} \label{sec:Review}
%
We express the exact unitary $U$ generated by the Hamiltonian $H(t)$ as a decomposition of an ideal unitary $U_{\mathrm{I}}$ and a noise unitary $\tilde{U}_{\mathrm{N}}$:
\beq \label{eqt:TotalU}
U(t, t_0) = \mathrm{T}\exp \left[ -i \int_{t_0}^{t} d\tau H(\tau) \right] = U_{\mathrm{I}}(t, t_0) \tilde{U}_{\mathrm{N}}(t, t_0) \ ,
\eeq
where $\mathrm{T}\exp$ denotes the time-ordered exponential and $U_{\mathrm{I}}(t,t_0)$ satisfies $\frac{d}{dt} U_{\mathrm{I}}(t,t_0) = -i H_{\mathrm{I}}(t) U_{\mathrm{I}}(t,t_0)$. Note that we assume units where $\hbar=1$. The equation of motion for $\tilde{U}_{\mathrm{N}}(t, t_0)$ is given by:
\begin{eqnarray}
\frac{d}{dt} \tilde{U}_{\mathrm{N}}(t,t_0) &=& -i U_{\mathrm{I}}(t,t_0)^\dagger H_{\mathrm{N}}(t) U_{\mathrm{I}}(t,t_0) \tilde{U}_{N}(t,t_0) \nonumber \\
&=& -i\tilde{H}_{\mathrm{N}}(t) \tilde{U}_{N}(t,t_0) \ . 
\end{eqnarray}
We can formally solve this using a Magnus expansion \cite{Magnus1954,Blanes2009},
\beq \label{eqt:tildeUN}
\tilde{U}_{\mathrm{N}}(t,t_0) = \exp \left[ -i \sum_{n=1}^{\infty} \tilde{\Phi}_n(t, t_0) \right] \ .
\eeq
The first-order term is given by:
\beq \label{eqt:1stMagnus}
\tilde{\Phi}_1(t, t_0) =  \int_{t_0}^{t} d\tau \tilde{H}_{\mathrm{N}}(\tau) \ . 
\eeq
The second-order term in our Magnus expansion is:
\beq
\tilde{\Phi}_2(t, t_0) = -\frac{i}{2}\int_{t_0}^{t} dt_1 \int_{t_0}^{t_1} dt_2 \ \left[\tilde{H}_{\mathrm{N}}(t_1), \tilde{H}_{\mathrm{N}}(t_2) \right] \ .
\eeq
We note that both $\tilde{\Phi}_1$ and $\tilde{\Phi}_2$ are Hermitian operators, so the operator $\exp(-i \left( \tilde{\Phi}_1(t,t_0) + \tilde{\Phi}_2(t,t_0) \right)$ is unitary.

To proceed, let us assume that 
\beq \label{eqt:Hnoise}
H_{\mathrm{N}}(t) = \sum_{\alpha=1}^n \eta_{\alpha}(t) B_\alpha(t) \ , 
\eeq
where $\eta_\alpha(t)$ is a pure real stochastic process and $B_{\alpha}(t)$ is a time-dependent Hermitian operator. We then have:
\begin{eqnarray} \label{eqt:tildeHN}
\tilde{H}_{\mathrm{N}}(t,t_0) &=& U_{\mathrm{I}}(t,t_0)^\dagger H_{\mathrm{N}}(t) U_{\mathrm{I}}(t,t_0) \nonumber \\
&=& \sum_{\alpha} \eta_{\alpha}(t) U_{\mathrm{I}}(t,t_0)^\dagger B_{\alpha} (t) U_{\mathrm{I}}(t,t_0) \nonumber \\
&=& \sum_{\alpha} \eta_{\alpha}(t) \tilde{B}_{\alpha}(t,t_0) \ , 
\end{eqnarray}
where we have defined the interaction picture noise operator $\tilde{B}_{\alpha}(t,t_0)  = U_{\mathrm{I}}(t,t_0)^\dagger B_{\alpha} (t) U_{\mathrm{I}}(t,t_0)$. In order to separate out the time-dependence of the operators, we assume that $\left\{ P_k \right\}$ form a Hermitian orthonormal basis of operators for the vector space of linear operators (Liouville space), and we can expand:
\beq \label{eqt:tildeB}
\tilde{B}_{\alpha}(t,t_0) = \sum_{k=1}^{d^2} \tilde{\mathcal{B}}_{\alpha k}(t,t_0) P_k \ ,
\eeq
where 
\begin{eqnarray} \label{eqt:Bsuperoperator}
\tilde{\mathcal{B}}_{\alpha k}(t,t_0) &=& \Tr \left( P_k \tilde{B}_{\alpha}(t,t_0) \right) \nonumber \\
&=& \Tr \left( P_k U_{\mathrm{I}}^\dagger(t,t_0) B_{\alpha}(t) U_{\mathrm{I}}(t,t_0)\right) \nonumber \\
&=& \Tr \left( B_{\alpha}(t) U_{\mathrm{I}}(t,t_0) P_k U_{\mathrm{I}}^\dagger(t,t_0) \right)  \ .
\end{eqnarray}
We then have for the first-order Magnus term:
\beq \label{eqt:1stOrder}
\tilde{\Phi}_1(t,t_0) =  \sum_{k=1}^{d^2}\left(\int^{t}_{t_0} d\tau \sum_{\alpha=1}^n \eta_{\alpha}(\tau) \tilde{\mB}_{\alpha k} (\tau, t_{0})  \right)  P_k \ .
\eeq

We can express $\tilde{\mathcal{B}}_{\alpha k}(t,t_0)$ compactly in the superoperator formalism as the inner product $\left( B_{\alpha}(t) | \mathcal{U}_{\mathrm{I}}(t,t_0) | P_k \right)$, where $\mathcal{U}_{\mathrm{I}}$ is the superoperator associated with $U_{\mathrm{I}}$: $\mathcal{U}_{\mathrm{I}}(t,t_0) = U_{\mathrm{I}}(t,t_0) \odot U_{\mathrm{I}}^\dagger(t,t_0)$, where $\odot$ denotes the linear operator argument of the superoperator. Furthermore, if we suppose that $t > t_1 > t_0$, then we can calculate $\tilde{\mathcal{B}}_{\alpha k}(t,t_0)$ from its value at time $t_1$ as follows:
\begin{eqnarray}
\tilde{\mathcal{B}}_{\alpha k}(t,t_0) &=& \Tr \left( B_{\alpha}(t) U_{\mathrm{I}}(t, t_1) U_{\mathrm{I}}(t_1,t_0) P_k \right. \nonumber \\
 && \left. U_{\mathrm{I}}^\dagger(t_1,t_0) U_{\mathrm{I}}^\dagger(t,t_1) \right) \nonumber \\
&=& \left( B_\alpha(t) | \mathcal{U}_{\mathrm{I}}(t,t_1) \mathcal{U}_{\mathrm{I}}(t_1,t_0) | P_k \right) \nonumber \\
&=& \sum_{l=1}^{d^2} \tilde{\mB}_{\alpha l}(t,t_1) \left(\mathcal{U}_{\mathrm{I}}(t_1,t_0) \right)_{lk} \ .
\end{eqnarray}

The derivation for the second-order Magnus term proceeds in a similar way.  Using our expressions in Eqs.~\eqref{eqt:tildeHN} and \eqref{eqt:tildeB}, we have:
\begin{eqnarray} \label{eqt:2ndOrder}
\tilde{\Phi}_2(t,t_0) &=& - \frac{i}{2} \sum_{k,k'=1}^{d^2}  \left( \int_{t_0}^t dt_1  \sum_{\alpha=1}^n \eta_{\alpha}(t_1) \tilde{\mB}_{\alpha k}(t_1,t_0) \right. \nonumber \\
&& \hspace{-1cm} \left. \int_{t_0}^{t_1} dt_2 \sum_{\alpha'=1}^n \eta_{\alpha'}(t_2) \tilde{\mB}_{\alpha' k'}(t_2,t_0) \right) \left[ P_k , P_{k'} \right] \ . \nonumber \\
\end{eqnarray}
It is now convenient to consider the superoperator associated with the unitary $\tilde{U}_{\mathrm{N}}$, given by
\beq
\tilde{\mU}_{\mathrm{N}}(t,t_0) = \tilde{U}_{\mathrm{N}}(t,t_0) \odot \tilde{U}_{\mathrm{N}}(t,t_0)^\dagger = e^{-i \sum_{n=1}^{\infty} \tilde{\mL}_n(t,t_0)} \ ,
\eeq
where $\tilde{\mL}_n(t,t_0)$ are the superoperator Magnus expansion terms:
\bes
\begin{align}
\tilde{\mL}_1(t,t_0) &= \left[ \tilde{\Phi}_1(t,t_0), \odot \right] \ , \label{eqt:1stMagnus} \\
\tilde{\mL}_2(t,t_0) &=  \left[ \tilde{\Phi}_2(t,t_0), \odot \right] \ .
\end{align}
\ees

In order to perform an ensemble average over stochastic noise realizations, which we denote by $\overline{\odot}$, we use the cumulant expansion.
The cumulant expansion of $\tilde{\mU}_{\mathrm{N}}$ is given by \cite{Kubo1962}:
\beq \label{eqt:cumulant}
\overline{\tilde{\mU}_{\mathrm{N}}(t,t_0)} = e^{\mK(t,t_0)} \ ,
\eeq
and
\beq \label{eqt:K}
\mK(t,t_0) = \sum_{k=1}^{\infty} \frac{(-i)^k}{k!} \overline{\left(\sum_{n=1}^{\infty} \tilde{\mL}_n(t,t_0) \right)^k}^c \ ,
\eeq
where $\overline{\odot}^c$ denotes the cumulant average. For example:
\beq \label{eqt:CumulantAverage}
\overline{\eta_{\alpha}(t)}^c = \overline{\eta_{\alpha}(t)} \ , \quad \overline{\eta_{\alpha}(t)\eta_{\beta}(t)}^c = \overline{\eta_{\alpha}(t)\eta_{\beta}(t)} - \overline{\eta_{\alpha}(t)} \cdot \overline{\eta_{\beta}(t)}  \ .
\eeq
If we consider the $k=1$ term of $\mK$, we have:
\beqnn
\overline{\left(\sum_{n=1}^{\infty} \tilde{\mL}_n(t,t_0) \right)}^c = \left(\sum_{n=1}^{\infty} \overline{\tilde{\mL}_n(t,t_0)} \right) = \left(\sum_{n=2}^{\infty} \overline{\tilde{\mL}_n(t,t_0)} \right) \ ,
\eeqnn
where we have assumed that the noise has mean zero so the average of the $n=1$ term gives zero.  Therefore, the leading-order behavior of $\mK$ is given by:
\begin{eqnarray} \label{eqt:LeadingOrderCumulant}
\mK(t,t_0) &=& -i \left( \overline{\tilde{\mL}_2(t,t_0) } + \dots \right) \nonumber \\
&& - \frac{1}{2} \left( \overline{\tilde{\mL}_1(t,t_0)^2} + \dots \right) + \dots \ , 
\end{eqnarray}
where the first (second) term arises from the $k=1$ ($k=2$) term, respectively.  We give explicit expressions for these two terms in Appendix~\ref{app:Cumulant}, which reproduce the results of Ref.~\cite{Hangleiter2021}.  The first term is anti-Hermitian, $\left( -i \overline{\tilde{\mL}_2(t,t_0) } \right)^\dagger = i \overline{\tilde{\mL}_2(t,t_0) }$, so it corresponds to coherent errors since it generates unitary dynamics. The second term is Hermitian,  $\left(\overline{\tilde{\mL}_1(t,t_0)^2}\right)^\dagger = \left(\overline{\tilde{\mL}_1(t,t_0)^2}\right)$, so it corresponds to decoherence described by a Liouvillian.  Therefore, the evolution is approximated up to second order in the noise strength by:
\beq 
\mathcal{U}(t,t_0) \approx \mathcal{U}_{\mathrm{I}}(t,t_0) e^{ -i  \overline{\tilde{\mL}_2(t,t_0) }  - \frac{1}{2}  \overline{\tilde{\mL}_1(t,t_0)^2}} \ .
\eeq

For a fixed $t$ and $t_0$, the error associated with truncating the Magnus and cumulant expansion can be reduced by truncating at a higher order.  However, another source of error can arise in calculating the integrals in Eqs.~\eqref{eqt:1stOrder} and \eqref{eqt:2ndOrder} numerically. If Eq.~\eqref{eqt:Bsuperoperator} must be computed on a discrete time grid, for example if $U_{\mathrm{I}}(t, t_0)$ is not known analytically as a function of $t$, then this time grid must be sufficiently fine such that these computations have an error that is smaller than that due to the truncated Magnus and cumulant expansion.
\subsection{A Different Ensemble Average} \label{sec:BridgeEnsembleAverage}
%
We now consider a different approach for performing the ensemble average in the cumulant expansion in Eq.~\eqref{eqt:cumulant}. We restrict to the case where the classical noise processes $\left\{ \eta_{\alpha}(t) \right\}$ in $H_{\mathrm{N}}$ (Eq.~\eqref{eqt:Hnoise}) are given by an arbitrary sum of OU processes:
\beq \label{eqt:sumofOU}
\eta_{\alpha}(t) = \sum_{n} X_{\alpha n}(t) \ ,
\eeq
where the set $\left\{ X_{\alpha n}(t)\right\}$ are independent OU processes. 

\subsubsection{Background}
%
An OU process $X(t)$ is a stochastic process satisfying the stochastic differential equation \cite{Uhlenbeck1930}
\beq
d X(t)= \gamma \left( \mu - X(t) dt \right)+ \sigma dW(t) \ ,
\eeq
where $\gamma, \sigma > 0$ and $\mu$ are parameters characterizing the process and $dW(t)$ denotes the Wiener increment. 
Given an initial value of $x_0$ at $t_0=0$, an analytical solution for the process is given by:
\begin{eqnarray} \label{eqt:OUformalsolMain}
\left( X(t) | X(t_0) = x_0 \right) &=&  \nonumber \\
&& \hspace{-3cm} x_0 e^{-\gamma (t-t_0)} + \mu \left( 1 - e^{-\gamma (t-t_0)} \right) \nonumber \\
&& \hspace{-3cm} + \frac{\sigma}{\sqrt{2 \gamma}} e^{-\gamma (t-t_0)} W(e^{2 \gamma (t-t_0)} -1) \ , \quad t \geq 0 \ ,
\end{eqnarray}
where $ \frac{1}{\sqrt{2 \gamma}} W(e^{2 \gamma t} - 1) = \int_0^t e^{\gamma s} d W(s)$ denotes a Wiener process~\cite{Wiener1923}. For completeness, we provide background information on OU processes in Appendix~\ref{app:OU}. In addition to their nice mathematical properties, sums of OU processes can approximate $1/f$ noise when their parameters are chosen appropriately \cite{Bernamont1937,Surdin1939,Dutta1981,Kaulakys2005,Paladino2014}. 

We now assume we have a realization of each OU process $X_{\alpha n} (t)$ at a discrete set of time increments $\left\{ t_0, t_1, \dots \right\}$ such that $X_{\alpha n} (t_k) = x_{\alpha n k}$. This is accomplished using the formula~\cite{Gillespie1996}:
\begin{eqnarray} \label{eqt:OUalgorithmMain}
x_{\alpha n k} &=& x_{\alpha n k-1} e^{-\gamma_{\alpha n}  \Delta t_k} + \mu_{\alpha n} \left( 1 - e^{-  \gamma_{\alpha n} \Delta t_k} \right) \nonumber \\
&& + \mathbf{n} \sqrt{\frac{\sigma_{\alpha n}^2}{2 \gamma_{\alpha n}} \left( 1 - e^{-2 \gamma_{\alpha n} \Delta t_k} \right)}  \ ,
\end{eqnarray}
where $ \mathbf{n} \sim \mathcal{N}(0, 1)$, $\Delta t_k = t_{k}-t_{k-1}$, and is exact for arbitrary step sizes $\Delta t_k$. 

Using these noise realizations, we consider a new OU process defined between $[t_{k-1}, t_k]$ that is conditioned on taking the specific values $x_{\alpha n k}$ and $x_{\alpha n k-1}$ at $t_k$ and $t_{k-1}$ respectively, and we denote this new process by $\eta_{\alpha n k}(t) \equiv \left( X_{\alpha n}(t) | X_{\alpha n}(t_{k-1})=x_{\alpha n k-1}, X_{\alpha n}(t_k)  = x_{\alpha n k} \right) $.  This process corresponds to an OU bridge process with boundary values $x_{\alpha n k-1}$ at $t_{k-1}$ and $x_{\alpha n k}$ at $t_k$.  From Eq.~\eqref{eqt:OUformalsolMain}, an analytical solution for the OU bridge process can be expressed for $t \in [t_{k-1},t_k]$ as:
\begin{eqnarray} \label{eqt:ConditionedOUMain}
\eta_{\alpha n k}(t) &=& 
x_{\alpha n k-1} e^{-\gamma_{\alpha n} (t-t_{k-1})} \nonumber \\
&& + \mu_{\alpha n} \left( 1 - e^{-\gamma_{\alpha n} (t-t_{k-1})} \right) \nonumber \\
&&  + \frac{\sigma_{\alpha n}}{\sqrt{2 \gamma_{\alpha n}}} e^{-\gamma_{\alpha n} (t-t_{k-1})} B(u_{\alpha n k}(t)) \ ,
\end{eqnarray}
where $u_{\alpha n k }( t) = e^{2 \gamma_{\alpha n} (t - t_{k-1})}-1$ and $B(u_{\alpha n k}(t)) = \left( W(u_{\alpha n k}(t)) |  X_{\alpha n}({t_{k-1})} = x_{\alpha n k-1} , X_{\alpha n}({t_k}) = x_{\alpha n k}\right)$ is the conditioned Wiener process.  The conditioned Wiener process can be modeled as:
\begin{eqnarray} \label{eqt:OUBridgeMain}
B(u_{\alpha n k}( t)) &=& \frac{u_{\alpha n k} (t)}{u_{\alpha n k} (t_k)} w_{\alpha n k} + \tilde{W}(u_{\alpha n k}( t)) \nonumber \\
&& \hspace{-1cm} - \frac{u_{\alpha n k}( t)}{u_{\alpha n k}({t_k})} \tilde{W}(u_{\alpha n k}({t_k})) \ , t \in [t_{k-1},t_k ] \ , \nonumber \\
\end{eqnarray}
where $\tilde{W}$ is an independent Wiener process and:
\begin{eqnarray}
w_{\alpha n k} &=& \frac{\sqrt{ 2 \gamma_{\alpha n}}}{\sigma_{\alpha n}}  \left[ e^{\gamma_{\alpha n} \Delta t_k}  x_{\alpha n k} - x_{\alpha n k-1}  \right. \nonumber \\
&& \left. - \mu_{\alpha n} \left( e^{\gamma_{\alpha n} \Delta t_k} - 1 \right)  \right] \ ,
\end{eqnarray}
The conditioned Wiener process $B(u_{\alpha n k}(t))$ satisfies $B({0}) = 0$ and $B(u_{\alpha n k}(t_k)) = w_{\alpha n k}$, which gives the desired boundary conditions for the OU bridge process in Eq.~\eqref{eqt:ConditionedOUMain}. We give a derivation of these expression in Appendix~\ref{app:OUBridge}.

We choose to express Eq.~\eqref{eqt:ConditionedOUMain} for the OU bridge process as a sum of two terms:
\begin{eqnarray} \label{eqt:OUBridgeProcessMain}
\eta_{\alpha n k}(t) &=& \eta^{(\mathrm{D})}_{\alpha nk}(t) + \eta^{(\mathrm{S})}_{\alpha n k } (t) \ .
\end{eqnarray}
The first term $\eta^{(\mathrm{D})}_{\alpha nk}(t)$ is defined as the expectation value of the OU bridge process:
\begin{eqnarray} 
\eta^{(\mathrm{D})}_{\alpha nk}(t) &=& \mathbb{E} \left[ \eta_{\alpha n k}(t) \right] \ ,  
\end{eqnarray} 
and it depends on the boundary values $x_{\alpha n k-1}, x_{\alpha n k}$ in a fixed way.  Specifically, for $\mu_{\alpha nk} = 0$ we have:
\begin{eqnarray} 
\eta^{(\mathrm{D})}_{\alpha nk}(t) &=&  x_{\alpha n k-1} \frac{\sinh \left(\gamma_{\alpha n} \left(t_k - t \right) \right)}{\sinh \left(\gamma_{\alpha n} \Delta t_k \right)} \nonumber \\
&& + x_{\alpha n k} \frac{\sinh(\gamma_{\alpha n} \left(t-t_{k-1} \right))}{\sinh(\gamma_{\alpha n}(\Delta t_k))}  \ .
\end{eqnarray} 
This follows from Eq.~\eqref{eqt:ConditionedOUMain}, and we give a derivation of this expression is given in Appendix~\ref{app:OUBridge}.

The second term in Eq.~\eqref{eqt:OUBridgeProcessMain}, $\eta^{(\mathrm{S})}_{\alpha n k } (t)$, is a zero-mean (stochastic) OU bridge process that evaluates to zero at $t_{k-1}$ and $t_{k}$ and is independent of the boundary values $x_{\alpha n k-1}, x_{\alpha n k}$ and hence the specific noise realization. This process is what we refer to as the zero-boundary bridge process, and it can be expressed as
\begin{eqnarray}
\eta^{(\mathrm{S})}_{\alpha n k } (t) &=&  \frac{\sigma_{\alpha n}}{\sqrt{2 \gamma_{\alpha n}}} e^{- \gamma_{\alpha n}(t - t_{k-1})} \nonumber \\
&& \hspace{-2cm} \times \left( \tilde{W}(u_{\alpha n k}(t)) - \frac{u_{\alpha n k}(t)}{u_{\alpha n k}(t_k)} \tilde{W}(u_{\alpha n k}(t_k)) \right)  , \nonumber \\
&& \text{for } t \in [t_{k-1},t_k ]  \ .
\end{eqnarray}
The stochastic process $\eta^{(\mathrm{S})}_{\alpha n k } (t)$ also describes an OU bridge process, but now where the process is fixed to be zero at $t_{k-1}, t_k$.  Furthermore, it does not depend on the specific realization $\lbrace x_{k-1}, x_{k} \rbrace$ of the coarse OU process. We can calculate the correlators of this process straightforwardly:
\bes  \label{eqt:BridgeOUMain}
\begin{align}
\overline{\eta_{\alpha n k}^{(\mathrm{S})}(t) }  & = 0 \ , \label{eqt:1stordercorrelation} \\
\overline{\eta_{\alpha n k}^{(\mathrm{S})}(t) \eta_{\alpha n k}^{(\mathrm{S})}(s) } &=  \frac{\sigma_{\alpha n}^2}{\gamma_{\alpha n}} \frac{ \sinh(\gamma_{\alpha n} (s - t_{k-1}))}{\sinh(\gamma_{\alpha n} \Delta t_k)} \nonumber \\
& \hspace{-1cm} \times  \sinh( \gamma_{\alpha n} (t_k-t)) \ ,  t_{k-1} \leq s \leq t \leq t_k \ . \label{eqt:2ndordercorrelation} 
\end{align}
\ees
Because we have assumed the OU processes are independent, the only non-zero two-point correlator involving $\eta^{(\mathrm{S})}_{\alpha n k } (t)$ is between the stochastic process with itself. Furthermore, by construction, the $\eta^{(\mathrm{S})}_{\alpha n k } (t)$  processes between different time increments are independent.  Therefore, we can write:
\beq \label{eqt:bridgeindependence}
\overline{\eta_{\alpha n k}^{(\mathrm{S})}(t) \eta_{\beta o k'}^{(\mathrm{S})}(s) } = \delta_{\alpha, \beta} \delta_{n,o} \delta_{k,k'} \overline{\eta_{\alpha n k}^{(\mathrm{S})}(t) \eta_{\alpha n k}^{(\mathrm{S})}(s) } \ .
\eeq 
%

\subsubsection{Two Step Ensemble Average}
%
We now repeat the analysis in Sec.~\ref{sec:Review} but where we condition each OU process in $H_{\mathrm{N}}(t)$ (Eqs.~\eqref{eqt:Hnoise} and \eqref{eqt:sumofOU}) on a noise realization of the stochastic process $X_{\alpha n}(t)$ specified on the time grid $\left\{t_0, t_1, \dots \right\}$. 
We summarize the steps of the derivation. First, Based on the time grid chosen for the noise realizations, we express our unitary $\mathcal{U}(t,t_0)$ as a sequence of unitaries, $ \prod_{k=1}^{} \mathcal{U}(t_k,t_{k-1})$, where for each $\mathcal{U}(t_k,t_{k-1}) =  \mathcal{U}_{\mathrm{I}}(t_k, t_{k-1}) { \tilde{\mathcal{U}}_{\mathrm{N}}(t_k, t_{k-1})}$ we perform the same Magnus expansion presented in Sec.~\ref{sec:Review}. 

Next, for each time interval $[t_{k},t_{k-1}]$, the conditioned noise Hamiltonian $H_{\mathrm{N}}$ (Eq.~\eqref{eqt:Hnoise}) can be expressed in terms of the conditioned stochastic processes $\eta_{\alpha n k}(t)$. Furthermore, we write $\eta_{\alpha n k}(t)$ in terms of a deterministic function $\eta_{\alpha n k}^{(\mathrm{D})}(t)$ and a zero-boundary bridge process $\eta_{\alpha n k}^{(\mathrm{S})}(t)$ as in Eq.~\eqref{eqt:OUBridgeProcessMain}. This allows us to express the first and second order Magnus terms as:
\bes \label{eqt:MagnusDS}
\begin{align}
\tilde{\mL}_1(t_k, t_{k-1}) & =  \tilde{\mL}_1^{(\mathrm{D})}(t_k, t_{k-1}) + \tilde{\mL}_1^{(\mathrm{S})}(t_k, t_{k-1}) \ , \\
\tilde{\mL}_2(t_k, t_{k-1}) & =  \tilde{\mL}_2^{(\mathrm{D})}(t_k, t_{k-1}) + \tilde{\mL}_2^{(\mathrm{S},\mathrm{D})}(t_k, t_{k-1}) \nonumber \\
&  + \tilde{\mL}_2^{(\mathrm{D},\mathrm{S})}(t_k, t_{k-1}) + \tilde{\mL}_2^{(\mathrm{S})}(t_k, t_{k-1}) \ ,
\end{align}
\ees
where $\tilde{\mL}_1^{(\mathrm{D})}(t_k, t_{k-1}),  \tilde{\mL}_2^{(\mathrm{D})}(t_k, t_{k-1}) $ only involve elements of $\left\{ \eta_{\alpha n k}^{(\mathrm{D})}(t) \right\}$ and their products, $\tilde{\mL}_1^{(\mathrm{S})}(t_k, t_{k-1}),  \tilde{\mL}_2^{(\mathrm{S})}(t_k, t_{k-1}) $ only involve elements of $\left\{ \eta_{\alpha n k}^{(\mathrm{S})}(t) \right\}$ and their products, and $ \tilde{\mL}_2^{(\mathrm{S},\mathrm{D})}(t_k, t_{k-1}),  \tilde{\mL}_2^{(\mathrm{D},\mathrm{S})}(t_k, t_{k-1})$ involve a product of a single element of $\left\{ \eta_{\alpha n k}^{(\mathrm{D})}(t) \right\}$ with a single element of $\left\{\eta_{\alpha n k}^{(\mathrm{S})}(t) \right\}$.

We then perform an ensemble average but only on the zero-boundary bridge processes $\left\{ \eta_{\alpha n k}^{(\mathrm{S})}(t) \right\}$. This ensemble average corresponds to averaging over all noise trajectories that connect the boundary terms $(x_{\alpha n k-1}, x_{\alpha n k})$,  as illustrated in Fig.~\ref{fig:SummaryFigure}(c). Because the zero-boundary bridge processes between different time intervals are independent (Eq.~\eqref{eqt:bridgeindependence}), we have the following property:
\begin{eqnarray} \label{eqt:concatenation}
\overline{ \prod_{k=1}^{}  \mathcal{U}_{\mathrm{I}}(t_k, t_{k-1}) \tilde{\mathcal{U}}_{\mathrm{N}}(t_k, t_{k-1})} &=& \nonumber \\
&& \hspace{-3cm} \prod_{k=1}^{}  \mathcal{U}_{\mathrm{I}}(t_k, t_{k-1}) \overline{ \tilde{\mathcal{U}}_{\mathrm{N}}(t_k, t_{k-1})} \ ,
\end{eqnarray}
which is also illustrated in Fig.~\ref{fig:SummaryFigure}(b). The cumulant expansion for each $\overline{ \tilde{\mathcal{U}}_{\mathrm{N}}(t_k, t_{k-1})}$ then takes the form:
\begin{eqnarray} \label{eqt:LeadingOrderCumulantBridge}
\mK(t_k,t_{k-1}) &=& \nonumber \\
&& \hspace{-2cm} -i \left( \tilde{\mL}_1^{(\mathrm{D})}(t_k,t_{k-1}) +  \tilde{\mL}_2^{(\mathrm{D})}(t_k,t_{k-1})  + \dots \right)  \nonumber \\
&& \hspace{-2cm} - \frac{1}{2}  \overline{\tilde{\mL}_1^{(\mathrm{S})}(t_k,t_{k-1})^2} - i \overline{\tilde{\mL}_2^{(\mathrm{S})}(t_k,t_{k-1})} + \dots  \ ,
\end{eqnarray}
where we have used that
 \beq
 \overline{\tilde{\mL}_1^{(\mathrm{S})}(t_k,t_{k-1})} = \overline{\tilde{\mL}_2^{(\mathrm{S,D})}(t_k,t_{k-1})} = \overline{\tilde{\mL}_2^{(\mathrm{D,S})}(t_k,t_{k-1})} = 0
 \eeq
 because $\overline{\eta_{\alpha n k}^{(\mathrm{S})}(t)} = 0$ (Eq.~\eqref{eqt:1stordercorrelation}). The second order terms in the cumulant expansion,
$\overline{\tilde{\mL}_1^{(\mathrm{S})}(t_k,t_{k-1})^2}, \overline{\tilde{\mL}_2^{(\mathrm{S})}(t_k,t_{k-1})}$, only depend on the 2-point correlations of $\left\{ \eta_{\alpha n k}^{(\mathrm{S})}(t) \right\}$ (Eq.~\eqref{eqt:2ndordercorrelation}). Further details to arrive at Eq.~\eqref{eqt:LeadingOrderCumulantBridge} are provided in Appendix~\ref{app:Cumulant}. 

We note that because $\eta^{(\mathrm{D})}_{\alpha n k}$ and the two-point correlation functions $\overline{\eta_{\alpha n k}^{(\mathrm{S})}(t) \eta_{\alpha n k}^{(\mathrm{S})}(s) }$ are known analytically (Eq.~\eqref{eqt:BridgeOUMain}), the integrals associated with the terms in Eq.~\eqref{eqt:LeadingOrderCumulantBridge} can be evaluated without the need of generating fine-grained noise realizations.

Similar to the results in the previous section, the superoperators $-i \tilde{\mL}_1^{(\mathrm{D})}(t_k,t_{k-1}),  -i \tilde{\mL}_2^{(\mathrm{D})}(t_k,t_{k-1})$, and $-i \overline{\tilde{\mL}_2^{(\mathrm{S})}(t_k,t_{k-1})}$ are Hermitian. The superoperator $ -\overline{\tilde{\mL}_1^{(\mathrm{S})}(t_k,t_{k-1})^2}$ is anti-Hermitian and corresponds to a Liouvillian with positive rates.  Therefore, truncating $\mathcal{K}$ at second order gives rise to a completely-positive trace-preserving (CPTP) quantum operation.

Therefore, the procedure for simulation proceeds as follows:
\begin{enumerate}
\item Generate a noise realization for each OU process $X_{\alpha n}(t)$ (Eq.~\eqref{eqt:sumofOU}) at some desired set of time points $\left\{t_0, t_1, t_2, \dots \right\}$.  Conditioning on this set of noise realizations, we can replace $X_{\alpha n}(t)$ in our Hamiltonian with $\eta^{(\mathrm{D})}_{\alpha nk}(t) + \eta^{(\mathrm{S})}_{\alpha n k } (t)$ for $t \in [t_{k-1}, t_k]$.
\item Evolve the density matrix through each time increment: $ \rho(t_k) = \mathcal{U}_\mathrm{I}(t_k, t_{k-1}) e^{\mathcal{K}(t_k,t_{k-1})} \rho(t_{k-1})$, where $\mathcal{K}$ (Eqt.~\eqref{eqt:LeadingOrderCumulantBridge}) depends on the specific noise realization generated in the previous step. This results in the evolution of a density matrix for a specific noise realization of the OU processes.
\item Repeat Steps 1 and 2 with new noise realizations to complete the ensemble average, as illustrated in Fig.~\ref{fig:SummaryFigure}(a)
\end{enumerate}
It is worth emphasizing that we choose to first evolve the state with $\mathcal{U}(t_N ,t_0) = \prod_{k=1}^N \mathcal{U}_\mathrm{I}(t_k, t_{k-1})e^{\mathcal{K}(t_k,t_{k-1})}$ for a specific coarse noise realization and then average the evolved state over different coarse noise realizations, as opposed to averaging the superoperator $\mathcal{U}(t_N ,t_0) $ over different coarse noise realizations. We choose the former approach since it allows us to capture the effect of slow drift processes as well as include intermediate measurements as our examples next illustrate. Assuming the cumulative truncation error after concatenation remains small, both approaches should approximate the standard ensemble average result  discussed in Sec.~\ref{sec:Review} over the range $[t_0, t_N]$.

We conclude by noting that we have the freedom to choose our time increments $\left\{t_1, t_2, \dots \right\}$ and that these increments need not be evenly spaced. For example, this can be convenient when coarse-graining over quantum operations that vary in duration, such as in a quantum circuit where single and two qubit gates may have different physical durations. Because the separation of the time increments dictates the duration of the bridge process, the separation effectively dictates which components of the noise are treated as decoherence versus coherent errors in a given increment. Using longer increments has the advantage of requiring fewer steps to realize the total dynamics, but it also incurs a higher approximation error in the truncated Magnus and cumulant expansion. We can therefore expect to use longer time increments as the noise strength becomes weaker without sacrificing accuracy.
%

\section{Results} \label{sec:Results}
%
\subsection{Single-Qubit Example} \label{sec:onequbit}
As a first example, we consider the case of a single-qubit Hamiltonian with only a noise term of the form $H(t) = \frac{1}{2} \eta(t) \sigma^x$, where $\sigma^x$ denotes the Pauli-$x$ operator and $\eta(t)$ is our classical stochastic process. In this case, the ideal evolution corresponds to the identity operation, and $B_\alpha(t) = \frac{1}{2} \sigma^x$. For simplicity, we consider the case where $\eta(t)$ is given by a single OU process, and we generate a realization of the process at time increments $\left\{t_k \right\}$.  We write the conditioned OU process in the interval $t \in [t_{k-1}, t_k]$ as $\eta^{(\mathrm{D})}_k(t)  + \eta^{(\mathrm{S})}_{k}(t)$, where the first term is deterministic and dependent on the realized noise trajectory and the second term is the zero-boundary bridge process that is independent of the realized noise trajectory. Because the noise operator $B_{\alpha}(t)$ commutes with $U_{\mathrm{I}}(t,t_0)$, we have a simple form for $\tilde{H}_{\mathrm{N}}(t) = \frac{1}{2} \eta(t) \sigma^x$. Because the time dependence is only in the stochastic process, we have that $\left[ \tilde{H}_{\mathrm{N}}(t_1), \tilde{H}_{\mathrm{N}}(t_2) \right] = 0$, so the terms $\tilde{\mathcal{L}}_2^{(\mathrm{D})}$ and $\overline{\tilde{\mathcal{L}}_2^{(\mathrm{S})}}$ vanish. We are therefore left with only contributions from $\tilde{\mathcal{L}}_1^{(\mathrm{D})}$ and $\overline{(\tilde{\mathcal{L}}_1^{(\mathrm{S})})^2}$. Our zero-boundary bridge-ensemble averaged quantum process then takes the form:
\begin{eqnarray}
\overline{\mathcal{U}(t_k,t_{k-1})} &=& \exp \left[ -i \int_{t_{k-1}}^{t_k} \eta^{(\mathrm{D})}_k(\tau) d \tau \left[ \frac{1}{2} \sigma^x , \odot \right] \right. \nonumber \\
&& \hspace{-2.5cm} \left.+ \iint_{t_{k-1}}^{t_k} \overline{\eta^{(\mathrm{S})}_{k}(\tau_1) \eta^{(\mathrm{S})}_{k}(\tau_2)} d\tau_1 d \tau_2 \frac{1}{4} \left( \sigma^x \odot \sigma^x - \odot \right)\right] \ , \nonumber \\
\end{eqnarray}
where the explicit forms of $\eta^{(\mathrm{D})}_{k}(\tau)$ and $\overline{\eta^{(\mathrm{S})}_{k}(\tau_1) \eta^{(\mathrm{S})}_{k}(\tau_2)}$ are given in Eqs.~\eqref{eqt:EConditionedOU} and \eqref{eqt:BridgeOU} respectively.  The first term in the exponential corresponds to a noise trajectory-dependent coherent error, with a strength given by the average drift of the noise over the time interval.  The second term corresponds to a noise trajectory-independent dephasing in the $\sigma^x$ basis, with a strength given by the two-point correlation of the zero-boundary bridge process.
%
\subsection{Two-Qubit Example} \label{sec:twoqubit}
As a second example, we consider the case of fluctuations of the exchange coupling between two spins:
\beq
H(t) = J ( 1 + \eta(t) ) \vec{S}_1 \cdot \vec{S}_2 
\eeq
where $S^a = \frac{1}{2} \sigma^a$ for $a = x,y,z$ are the spin-1/2 operators. 

Because we are only considering a single noise operator, we drop the $\alpha$ index. This is another example where the noise operator $B(t)$ commutes with $U_{\mathrm{I}}(t,t_0)$, and $\tilde{H}_{\mathrm{N}}(t)$ takes the simple form $\tilde{H}_{\mathrm{N}}(t) = \eta(t) J \vec{S}_1 \cdot \vec{S}_2$.  It then follows that  $ \overline{\tilde{\mL}^{(\mathrm{S})}_2(t_k,t_{k-1})} =  \tilde{\mathcal{L}}_2^{(\mathrm{D})}(t_k,t_{k-1})= 0$ as in our previous example, and we can calculate $ \overline{\tilde{\mL}^{(\mathrm{S})}_1(t_k,t_{k-1})^2}$ and $\tilde{\mathcal{L}}_1^{(\mathrm{D})}(t_k,t_{k-1})$ in a similar manner. If we consider the case of a single OU process and condition it on a specific realization at the time points $\left\{t_k \right\}$ like in our single-qubit example from the previous section, we find the integration over the zero-boundary bridge process gives:
\begin{eqnarray} 
 \overline{\tilde{\mL}^{(\mathrm{S})}_2(t_k,t_{k-1})} &=& 0  \ ,\\
-\frac{1}{2} \overline{\tilde{\mL}^{(\mathrm{S})}_1(t_k,t_{k-1})^2} 
&=&\iint_{t_{k-1}}^{t_k} dt_1 dt_2 \ \overline{\eta^{(\mathrm{S})}(t_1) \eta^{(\mathrm{S})}(t_2)}  \nonumber \\
&& \hspace{-2cm}\times   \left( L \odot L^\dagger - \frac{1}{2} \left\{ L^\dagger L, \odot \right\} \right) \ , \label{eqt:2QubitDephasing}
\end{eqnarray}
with $L = J \vec{S}_1 \cdot \vec{S}_2$.  We show in Appendix~\ref{app:2qOperatorBasis} how we can recover this  form using the formalism presented in Sec.~\ref{sec:Method}. Therefore, the integrated stochastic  process gives rise to dephasing in the eigenbasis of the operator $\vec{S}_1 \cdot \vec{S}_2$, corresponding to losing any coherence between the $S=0$ and $S=1$ subspaces.

Similarly, the deterministic contributions are given by:
\begin{eqnarray}
\tilde{\mathcal{L}}_1^{(\mathrm{D})}(t_k,t_{k-1}) &=& \int_{t_{k-1}}^{t_k} d\tau \eta_{\alpha}^{(\mathrm{D})}(\tau) \left[ J \vec{S}_1 \cdot \vec{S}_2, \odot \right]  \ , \nonumber \\ \\
\tilde{\mathcal{L}}_2^{(\mathrm{D})}(t_k,t_{k-1}) &=& 0 \ ,
\end{eqnarray}
corresponding to a constant shift to the  Hamiltonian that depends on the specific boundary values, which then results in a coherent over- or under-rotation.
\subsection{Full Permutation Dynamical Decoupling for an Exchange-only Qubit} \label{sec:EXonly}

For our first non-trivial example, we choose to simulate the full permutation dynamical decoupling (DD) protocol~\cite{Sun2024} for a 3-spin exchange-only qubit~\cite{Bacon2000,Kempe2001,Kempe2001b}. A basis for the 8-dimensional Hilbert space of the three spins can be expressed in terms of the simultaneous eigenstates of the commuting operators (1) $\vec{S} \cdot \vec{S}$, where $\vec{S} = S_x \hat{x} + S_y \hat{y} + S_z \hat{y}$ and $S_\alpha = \frac{1}{2} \left( {\sigma}_\alpha^{(1)} + {\sigma}_\alpha^{(2)} + {\sigma}_\alpha^{(3)} \right)$.  The eigenvalues of $\vec{S} \cdot \vec{S}$ are given by $S(S+1)$; (2) $\vec{S}_{12} \cdot \vec{S}_{12}$, where $\vec{S}_{12}$ is the total angular momentum of only the 1st and 2nd spin; (3) $S_z$, the total spin in the $z$-direction.  
Denoting the basis elements by $\ket{S,S_{12}, S_z}$, the computational basis of the encoded 3-spin exchange-only qubit is given by the $S = 1/2$ states:
\beq
\ket{0} \equiv \ket{1/2, 0, S_z} \ , \ \ket{1} \equiv \ket{1/2, 1, S_z} \ , 
\eeq
with $S_z \in \left\{-1/2, 1/2 \right\}$, while the remaining $S = 3/2$ states ($\ket{3/2, 1, -3/2}$, $\ket{3/2, 1, -1/2}$, $\ket{3/2, 1, 1/2}$, $\ket{3/2,1, 3/2}$) are leakage states.
Assuming an ideal Hamiltonian for the system of three spins of the form:
\beq
H_{\mathrm{I}}(t) = g \mu_{\mathrm{B}} B \sum_{i=1}^3 S_i^z + \sum_{i=1}^2 J_{i,i+1}(t) \vec{S}_i \cdot \vec{S}_{i+1} \ , 
\eeq
pulsing $J_{12}$ or $J_{23}$ preserves the $S$ and $S_z$ quantum numbers of the state and allows for universal control of the encoded qubit~\cite{Bacon2000,Kempe2001,Kempe2001b}. Specifically, pulsing $J_{12}$ implements a rotation about the $-\hat{z}$ axis of the Bloch sphere of the qubit, whereas pulsing $J_{23}$ implements a rotation about the axis $\hat{n} = \left( \sqrt{3}/2, 0, 1/2 \right)$. We assume a global magnetic field of $B = 50\mu$T.

The full permutation DD protocol~\cite{Sun2024} proceeds as follows. A $\pi$ rotation about $\hat{n}$ followed by a $\pi$ rotation about $-\hat{z}$ is implemented three times sequentially to generate a logical identity operation. If we choose a noise Hamiltonian of the form:
\begin{eqnarray} \label{eqt:EXNoise}
H_{\mathrm{N}}(t) &=& \sum_{i=1}^3 \delta B_i(t) S_i^{z}+  \sum_{i=1}^2 \delta J_{i,i+1}(t) \vec{S}_{i} \cdot \vec{S}_{i+1}  \ , \nonumber \\
\end{eqnarray}
then the effect of the noise at first order in the Magnus expansion is to implement a $y$-rotation on the qubit, where the rotation rate is determined by the magnetic field gradient between the spins generated by the magnetic noise $\delta B_i(t)$~\cite{Sun2024}. This has the important consequence that a qubit prepared along the $\hat{z}$ axis of the Bloch sphere will exhibit oscillations whereas a qubit prepared along the $\hat{y}$ axis will not.
 
For simplicity, we assume that the $\delta B_i$ are described by a sum of identical and independent zero-mean OU processes. Furthermore, we take $\delta J_{i,i+1}(t) = J_{i,i+1}(t) \xi_{i,i+1}(t)$, with $\xi_{i,i+1}$ given by a sum of identical and independent zero-mean OU processes. The parameters of the OU processes are chosen to fit identical free induction decay and exchange decay $T_2^\ast$ measurements of exchange-only qubits in Si/SiGe quantum dots~\cite{Weinstein2023} with a $1/f$ power spectral density (PSD). This tuning is described in Appendix~\ref{app:NoiseParameterTuning}, where we also show excellent agreement between simulations, using our method, of free induction decay and exchange decay experiments and the chosen $T_2^\ast$ values.

For our simulations of the DD protocol, we assume square pulses of $10$ns duration for $J_{12}$ and $J_{23}$, each followed with a $10$ns idle corresponding to $J_{12} = J_{23} = 0$. We consider a coarse graining that covers the entire DD sequence, meaning our coarse noise realizations have a time step of $120$ns.

In the numerical simulations that follow, since the system Hilbert space of dimension $2^n$ is given by a tensor product of $n$ 2-dimensional Hilbert spaces, we choose the the Pauli operator basis as our orthonormal basis $\left\{P_k \right\}$:
\beq
P_{ \sum_{i=0}^{n-1} k_i 4^i} = \frac{1}{2^{n}} \sigma_{k_0} \otimes  \sigma_{k_1} \otimes \dots \otimes \sigma_{k_{n-1}} \ ,  k_{i} \in \left\{ 0,1,2,3 \right\} \ ,
\eeq
where $\sigma_0 = \ident$ and $\sigma_{1,2,3}$ are the 2-dimensional Pauli operators.
We choose to represent the superoperator $\mathcal{K}$ as a Pauli Transfer Matrix of dimension $4^n \times 4^n$, where its matrix elements are given by:
\beq
\mathcal{K}_{ij} = \mathrm{Tr} \left( P_i \mathcal{K}(P_j)\right) \ .
\eeq
The density matrix $\rho$ is then represented as a vector of dimension $4^n$ with elements:
\beq
\rho_i =  \mathrm{Tr} \left( P_i \rho \right) 
\eeq
\begin{figure}[!tb] 
   \centering
   \subfigure[]{\includegraphics[width=3.25in]{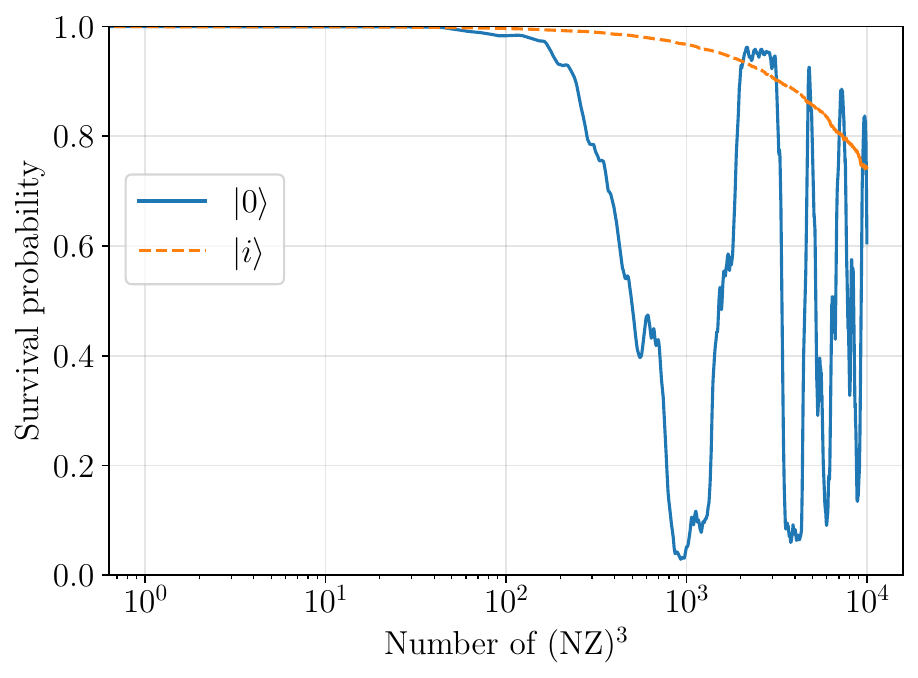} \label{fig:NZ3individual}}
   \subfigure[]{\includegraphics[width=3.25in]{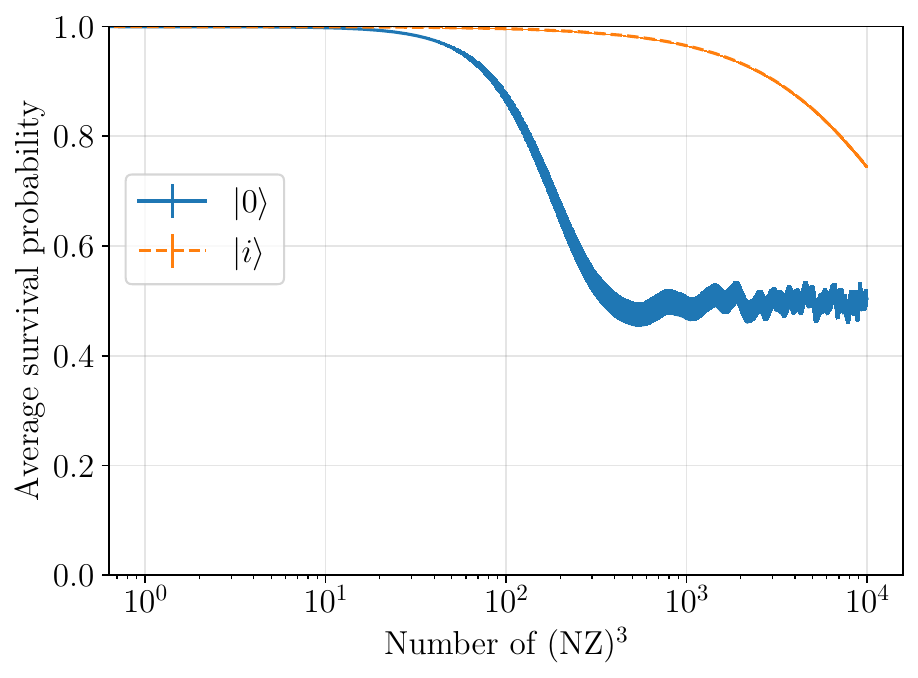}  \label{fig:NZ3ensemble}}
   \caption{Survival probability as a function of the number of applications of the DD sequence, which we denote by (NZ)$^3$, for two initial states for the exchange-only qubit.  The states $\ket{0}, \ket{i} = (\ket{0} + i \ket{1})/\sqrt{2}$ correspond to qubit states along the $z$ axis (blue solid curve) and $y$ axis (orange dashed curve) of the Bloch sphere respectively.  The survival probability in (a) is for a single coarse noise realization, while (b) is for the ensemble average over $10^3$ independent coarse noise realizations, with the error bars being the $2 \sigma$ confidence interval as estimated by performing a bootstrap over the independent simulations.}
   \label{fig:NZ3}
\end{figure}

In Fig.~\ref{fig:NZ3individual}, we show the survival probability calculated using our simulation method for two different initial states using an identical coarse noise realization as a function of the number of complete DD sequences applied (a complete DD sequence involves a total of 6 exchange pulses). This corresponds to performing steps (1) and (2) described in the summary at the end of Sec.~\ref{sec:BridgeEnsembleAverage}. The survival probability is calculated by evolving the initial density matrix after  $N$ applications of the DD sequence and calculating the probability of measuring the initial state. 
We see that the survival probability for the state initialized in the encoded $\ket{0}$ state (along the $z$-axis of the qubit Bloch sphere) exhibits oscillations, which from the analysis of Ref.~\cite{Sun2024} can be attributed to the $y$ rotation arising from the magnetic field gradients between the spins, which in our noise model (Eq.~\eqref{eqt:EXNoise}) are induced by the magnetic noise.  Because the magnetic noise is time-dependent, the frequency of these oscillations can vary as the number of applications of the DD sequence increases.  In contrast, the $y$ initial state (corresponding to the encoded state $\ket{i} = (\ket{0} + i \ket{1})/\sqrt{2}$) does not exhibit these oscillations since it should remain invariant (up to a global phase) under a $y$ rotation.  It however exhibits the expected decay in survival probability associated with decoherence, which is induced by the ensemble average over the zero-boundary bridge processes.

In Fig.~\ref{fig:NZ3ensemble}, we show the survival probability for the two different initial states after we perform the ensemble average over the coarse noise realizations.  This corresponds to step (3) in the summary at the end of Sec.~\ref{sec:BridgeEnsembleAverage}. We see that the two initial states exhibit different decay curves because of the effect of the $y$ rotation induced by the noise, as observed in Ref.~\cite{Sun2024}.

We emphasize that the different behavior observed for the two initial states in Fig.~\ref{fig:NZ3} is the outcome of having a non-zero contribution from the term $\tilde{\mL}_1^{(\mathrm{D})}$, which arises from the first order Magnus expansion term. Because we are considering zero-mean OU processes for our noise (Eq.~\eqref{eqt:EXNoise}), the average magnetic field gradient induced by the noise is zero, but every run of the experiment does experience a non-zero magnetic field gradient on the spins, which affects the statistics of the survival probability. Our approach of first performing the ensemble average over the zero-boundary bridge processes, evolving the state, then performing the ensemble average over coarse noise realizations preserves this effect even when we perform our temporal coarse graining over a single DD sequence and concatenate the sequence as in our simulations. In order to observe the same behavior with the standard ensemble average discussed in Sec.~\ref{sec:Review}, we would need to to perform the ensemble average over the entire time evolution as opposed to a single DD sequence in order to preserve the noise correlations between the different applications of the DD sequence. 
\subsection{Weight-2 Parity Check using Three Singlet-Triplet Qubits} \label{sec:Weight2Parity}
\begin{figure}
 \centering
    \includegraphics[width=2.25in]{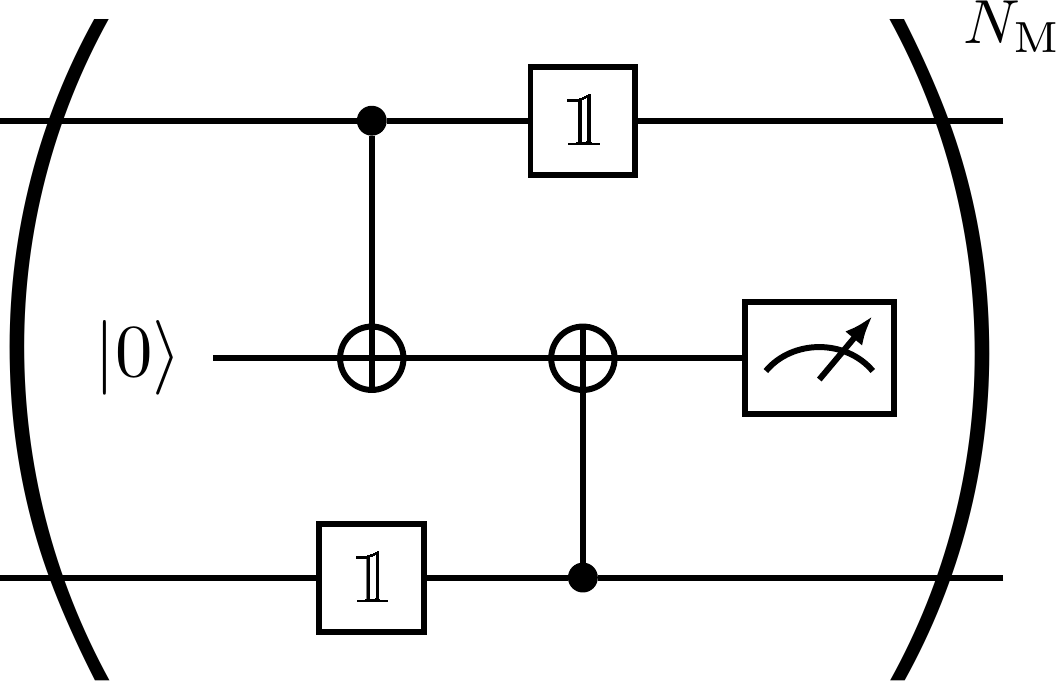} 
\caption{A weight-2 parity check circuit on two qubits (top and bottom wires) using a single ancilla qubit (center wire) prepared in the $\ket{0}$ state. The circuit is repeated a total of $N_{\mathrm{M}}$ times, which is represented by the large parentheses with the exponent $N_{\mathrm{M}}$.} \label{fig:weight2parity}
\end{figure}

As another non-trivial example, we consider a weight-2 parity check circuit depicted in Fig.~\ref{fig:weight2parity}. This example is meant to highlight one of the advantages of our approach, which is that it can handle simulating a sequence of mid-circuit measurements without additional overhead. For example, there are two approaches we could take to simulate a sequence of $N_{\mathrm{M}}$ mid-circuit measurements. The first approach, amenable to the Filter Function formalism, would be to fix the $N_{\mathrm{M}}$ measurement channels associated with specific outcomes and calculate the probability of observing such a sequence. This would require us to perform a total $2^{N_{\mathrm{M}}}$ simulations in order to calculate the probability of each of the possible sequences of measurement outcomes, which would be prohibitive if $N_{\mathrm{M}}$ is large. 

The second approach is to perform importance sampling~\cite{Clark1961,Tokdar2010,Elvira2021} by evolving the system up to each measurement and sample the measurement outcomes from the associated probability distribution. This approach is significantly more efficient when the number of independent simulations needed to build statistical confidence in estimates is significantly smaller than $2^{N_{\mathrm{M}}}$. However, if we were to perform such a simulation in the Filter Function formalism, calculating the outcome probabilities for any given measurement requires calculating the evolution of the state from its initial state and not simply from the previous measurement. This is the only way to propagate the noise correlations through the measurements. This is in contrast to our hybrid approach, where the state can be projected by the measurement and evolved to the next measurement because the noise correlations are carried naturally through the measurement by the coarse-grained noise trajectory. 

We choose to consider singlet-triplet qubits~\cite{Levy2002,Petta2005}, whereby a single qubit is encoded into the zero-magnetization states of two spins. The computational basis states are defined as the zero-magnetization singlet and triplet states:
\begin{eqnarray}
\ket{0} &\equiv&  \ket{S} \ , 
\ket{1} \equiv \ket{T_0} \ ,
\end{eqnarray}
The remaining two triplet states with non-zero magnetization are leakage states. The \emph{qubit} Pauli operators can then be identified in terms of the spin operators as~\cite{Bukard2023}:
\begin{eqnarray}
\sigma^x &\rightarrow& S_1^z - S_2^z \ ,  \nonumber \\
\sigma^y &\rightarrow& 2 \hat{z} \cdot \left( \vec{S_2} \times \vec{S_1} \right) \ , \nonumber \\
\sigma^z &\rightarrow& 2 \left(S_1^z S_2^z - \vec{S}_1 \cdot \vec{S}_2 \right) \ .
\end{eqnarray}

For the three-qubit problem (6 spins) in Fig.~\ref{fig:weight2parity}, we assume the spins are arranged in a line and are described by the ideal Hamiltonian:
\beq \label{eqt:6SpinH}
H_{\mathrm{I}}(t) = \sum_{i=1}^5 J_{i,i+1}(t) \vec{S}_i \cdot \vec{S}_{i+1}  + \mu_{\mathrm{B}} \sum_{i=1}^6 g_i {B}_i^z {S}_i^z \ . 
\eeq
Single-qubit gates on the three qubits are enacted by controlling the exchange interactions $J_{12}, J_{34}, J_{56}$ respectively, and two qubit gates require additionally controlling $J_{23}, J_{45}$.  In Appendix~\ref{app:SingletTriplet}, we present implementations of a single-qubit identity operation\footnote{Even in the absence of exchange interactions, the qubit Hamiltonian is given by $H = \mu_{\mathrm{B}} (g_i B_i - g_{i+1} B_{i+1}) \sigma^x$, which implements a constant-rate rotation around the $x$-axis of the Bloch sphere.  In our approach, we implement the identity operation by pulsing $J_{i,i+1}$ as opposed to waiting for the qubit to perform a full rotation around the Bloch sphere~\cite{Cerfontaine2014}.} and two-qubit CNOT gates for a particular choice of the magnetic fields $\left\{ g_i B_i^z \right\}$, where we assume that the exchange interaction pulses are given by $20$ ns square pulses followed by $20$ ns idle times. For our parameter choices, each CNOT gate takes $360$ ns.

For our noise Hamiltonian, we include magnetic noise along all three directions for each spin, and we include fluctuations for each exchange interaction.  The noise Hamiltonian is expressed as:
\begin{eqnarray} \label{eqt:ParityNoise}
H_{\mathrm{N}}(t) &=& \sum_{i=1}^6 \sum_{\alpha = x,y,z} \delta B_i^{\alpha}(t) S_i^{\alpha} \nonumber \\
&& + \sum_{i=1}^5  \delta J_{i,i+1}(t) \vec{S}_{i} \cdot \vec{S}_{i+1} \ .
\end{eqnarray}
For simplicity, we assume that the $\delta B_i^{\alpha}$ are described by identical and independent stochastic processes. Furthermore, we take $\delta J_{i,i+1}(t) = J_{i,i+1}(t) \xi_{i,i+1}(t)$.

In what follows, we study the behavior of the parity measurement outcomes under different noise processes. Specifically, we choose:
\begin{enumerate}
\item $1/f$ Noise Model: The stochastic processes $\delta B_i^{\alpha}$ and $\xi_{i,i+1}$ are given by a sum of independent OU processes such that the PSD of each is approximately $1/f$ in the range between $f_{\min}$ and $f_{\max}$.  This is achieved by having the $\gamma$ parameter of the OU processes be linearly spaced on a log scale between $2 \pi f_{\min}$ and $2 \pi f_{\max}$, and for each OU process we take $\sigma^2 = p \gamma$. This is the Gaussian noise analogue of generating approximately $1/f$ noise from an ensemble of independent telegraph noise sources, each of which contributes an OU-like Lorentzian power spectral density \cite{Dutta1981}.  Therefore, the two sets of $\left\{\delta B_i^{\alpha} \right\}$ and $\left\{ \xi_{i,i+1} \right\}$ are characterized by (a) an $f_{\min}$ and $f_{\max}$, (b) the number of OU processes, and (c) the parameter $p$. 
\item Quasi-static Noise Model: The stochastic processes $\delta B_i^{\alpha}$ and $\xi_{i,i+1}$ are given by quasi-static noise. This corresponds to taking $\sigma^2 = p \gamma$ followed by $\gamma \to 0$ for an OU process. Additional details about how quasi-static noise is a limit of the OU process are given in Appendix~\ref{app:OtherNoise}. This will mean that noise processes take a constant value during our simulations (this means $\tilde{\mathcal{L}}_{1,2}^{(\mathrm{S})} = 0$, because the bridge process is identically zero), but where the value is a Gaussian random number with variance $p/2$.  Therefore, the two sets of $\left\{\delta B_i^{\alpha} \right\}$ and $\left\{ \xi_{i,i+1} \right\}$ are characterized by a single parameter $p$. 

\item Bernoulli Noise Model: A useful example to contrast against our physical models above is the case where the parity-flip rate is constant and independent of the parity sector.  In this case, the time series of parity flips is described by a Bernoulli process. (The measurement outcomes are described by a telegraph process with a constant rate.) This noise model is parameterized by a single parameter $q$ corresponding to the probability of a parity flip at any given measurement.  This model does not require any quantum system evolution to be simulated and can be simulated by performing $N_{\mathrm{M}}$ independent Bernoulli trials.
\end{enumerate}

For the $1/f$ noise model, we use the same parameters as those in Sec.~\ref{sec:EXonly}. We tune the parameters of the quasi-static noise model so that it gives the same identical free induction decay and exchange decay $T_2^\ast$ times as the $1/f$ noise model, even though the functional dependence of the decay can vary from $e^{-t/T_2^\ast}$ to $e^{-(t/T_2^\ast)^3}$. This tuning is described in Appendix~\ref{app:NoiseParameterTuning}.

We now proceed to simulate repeated weight-2 parity check measurements. We use the same Pauli operator basis as in Sec.~\ref{sec:EXonly}. We consider a coarse graining that covers a single pulse plus idle, meaning our coarse noise realizations have a time step of $40$ns. We show in Appendix~\ref{app:SingletTriplet} that we get the same results if we coarse grained over multiple pulses where the coarse grained time step is $120$ns. For each coarse noise realization generated, we evolve the density matrix up to the measurement.  For simplicity, we assume the measurement is instantaneous and error free, where the measurement outcomes correspond to measuring the singlet or any of the three triplet states for the two-spin system. We calculate the measurement outcome probabilities from the evolved density matrix, and we project onto one of the measurement outcomes randomly according to the probabilities.
We also assume the reset of the auxiliary qubit is instantaneous and error-free.  The OU processes that comprise $\delta B_i^{\alpha}$ and $\delta J_{i, i+1}$ in Eq.~\eqref{eqt:ParityNoise} are continued through the measurements so that we can study the role of drift on the measurement outcomes.  This is repeated a total of $N_{\mathrm{M}}$ times, so that for each coarse noise realization we have a $N_{\mathrm{M}}$ bitstring corresponding to the measurement outcomes.  We can then study properties of the distribution of bitstrings as discussed next.

\begin{figure}[!t] %
   \centering
    \includegraphics[width=3.25in]{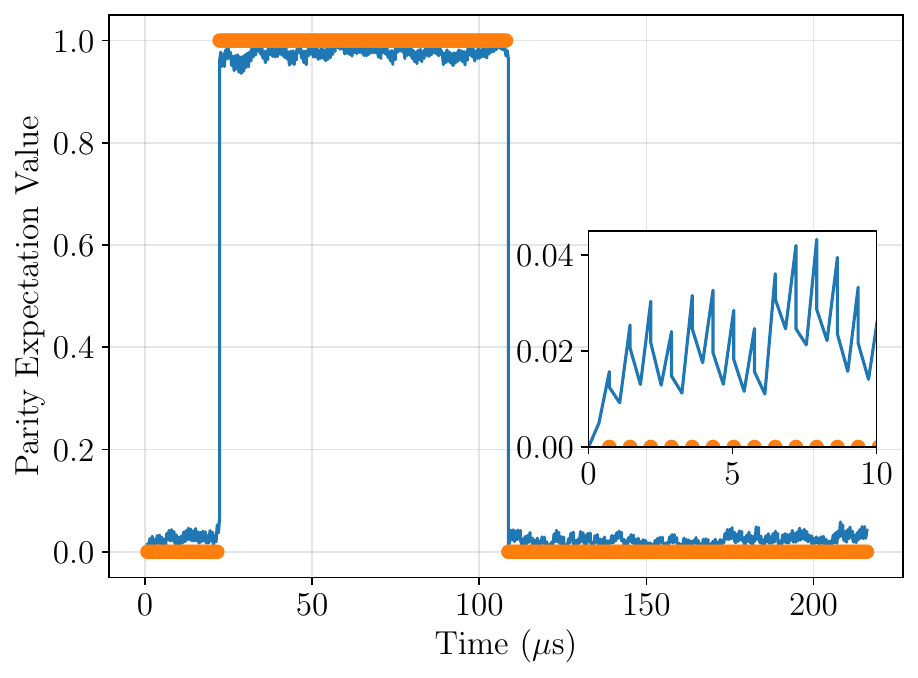} 
   \caption{A single realization of the parity expectation value of the data qubits $\langle \frac{1}{2} \left( \ident - \sigma_1^z \sigma_3^z \right) \rangle$ (solid blue line) and the measurement outcomes (orange circles) for the $1/f$ noise model with no idle time between measurements. A total of 300 measurements are shown, and each measurement is separated by $720$ns corresponding to the time required to implement the two CNOTs. Inset: Zoom of the early time behavior where the parity expectation value is close to 0.}
   \label{fig:PartiyExp}
\end{figure}

In order to understand the role of the finite-time CNOTs in causing parity flips, we first consider the case where the CNOTs are instantaneous so that there is no time for errors to accumulate on the data and measurement qubits.  In this case, we expect the measurement outcome to always be 0 (even parity) irrespective of how many times the parity measurement is performed.  We now consider the effect of finite-time CNOTs, and we show in Fig.~\ref{fig:PartiyExp} one realization of the measurement outcomes for a sequence of 300 measurements, where we plot the parity expectation value of the data qubits $\langle \frac{1}{2} \left( \ident - \sigma_1^z \sigma_3^z \right) \rangle$ as well as the measurement outcomes. We can make two observations about the effect of the imperfect CNOTs.  First, we observe that the accumulation of errors during the implementation of the CNOTs can result (although with small probability for our parameter choices) in parity measurement errors and the parity of the data qubits flipping. Second, the act of performing the parity measurement does not perfectly project the data qubits to the parity sector associated with the measurement outcome. This is most easily seen in the inset of Fig.~\ref{fig:PartiyExp}, where the parity expectation value is not restored to 0 even though a 0 is measured. Therefore, there is the additional possibility of residual errors in the data qubits corresponding to not being in a single parity sector even after a measurement is performed.
\begin{figure}[!t] %
   \centering
   \includegraphics[width=3.25in]{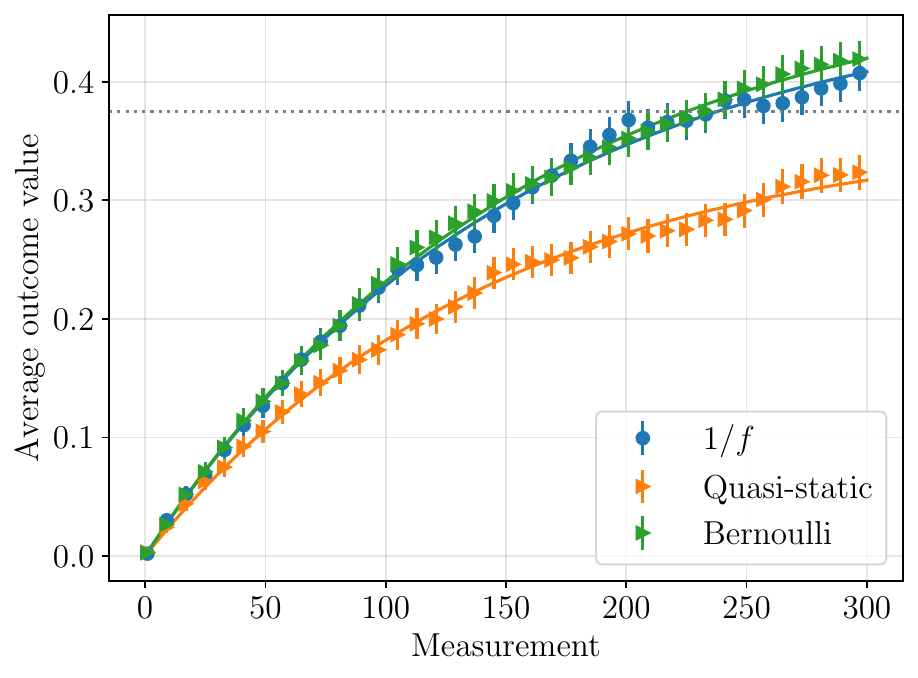} 
   \caption{The expectation value of the measurement outcome during the sequence of weight-2 parity measurements for different noise models. The Bernoulli process uses a parity-flip probability of $q=3 \times 10^{-3}$ in order to closely match the results of the $1/f$ noise model.  The solid lines correspond to fits to $a \left( 1 - e^{2 \lambda t} \right)$ with $(a, \lambda) = (0.475 \pm 0.004, 0.00327 \pm 6 \times 10^{-5}), (0.361 \pm 0.004, 0.00351 \pm 8 \times 10^{-5}), (0.494 \pm 6 \times 10^{-3}, 0.00316 \pm 6 \times 10^{-5})$ for the $1/f$, quasi-static and Bernoulli processes. The dotted line corresponds to the value of 3/8. The results are averaged over a total of $4 \times 10^3$ independent simulations, with the error bars being the $2 \sigma$ confidence interval as estimated by performing a bootstrap over the independent simulations. We only show every 8-th measurement for clarity.} \label{fig:OutcomeExp}
\end{figure}

When averaged over independent noise realizations, we can study the average measurement outcome for each measurement. For the early measurements, we expect the average to be close to 0 since the data qubits are unlikely to have accumulated enough errors to cause a parity flip.  As the number of CNOTs and measurements performed increases, we expect this value to grow and to saturate at 1/2 in the absence of leakage, whereby the system has equal probability of being measured in either parity sector. In the presence of leakage, the saturation value may vary.  For example, if each of the four states of each data qubit are equally likely, then the expectation value should be $3/8$ in the absence of any other source of error\footnote{Because all triplet states are measured as `1', there are a total of ten basis states that give rise to 0 measurement and six basis states that give rise to a 1 measurement.}.

We show our simulation results for the average measurement outcome in Fig.~\ref{fig:OutcomeExp}, where the probability of a parity flip for the Bernoulli model is chosen to best approximate the behavior of the $1/f$ model. We find that although the $1/f$ and quasi-static noise models were tuned to have the same $T_2^\ast$ times, they exhibit quantitatively different features.  In addition to having different parity-flip rates, the quasi-static model saturates to a value closer to 3/8 compared to 1/2 for the $1/f$ model, suggesting that leakage is more pronounced for the quasi-static model.

We find that the expectation value of measurement outcomes at each measurement step (Fig.~\ref{fig:OutcomeExp}) can be well fit to the function $a \left( 1- e^{-2 \lambda t} \right)$. This is the expected functional form for the behavior of the Bernoulli parity-flip process starting in the even parity sector with a constant transition rate $\lambda$ for both transitions, but we find that the $1/f$ and quasi-static noise models also fit this behavior very well.  However, as we show next this does not mean that these two noise models have a constant transition rate.

\begin{figure}[!t] %
   \centering
    \includegraphics[width=3.25in]{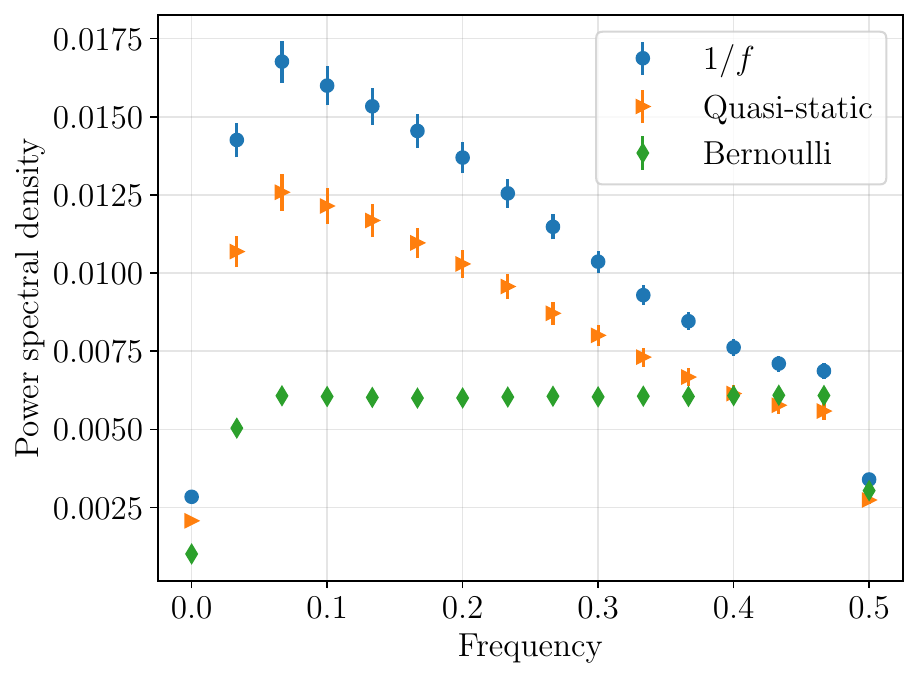} 
   \caption{PSD of the parity-flip time series (Eq.~\eqref{eqt:parityflipTS}) derived from the measurement outcomes for the two simulated noise models and an independent simulation of a Bernoulli process. The simulated Bernoulli process uses a probability of $q=3 \times 10^{-3}$ and 300 measurements in order to match the other simulations. The PSD is estimated using Welch's method \cite{Welch1967} as implemented in SciPy 1.14.1 \cite{SciPy} using a segment length of 30.  The results are averaged over a total of $4 \times 10^3$ independent simulations, with the error bars  (mostly hidden by the markers) being the $2 \sigma$ confidence interval as inferred by performing a bootstrap over each simulation's estimated PSD. }
   \label{fig:PartiyPSD}
\end{figure}

In order to study the statistics of parity-flip events, we define the time series of parity-flip events by:
\beq \label{eqt:parityflipTS}
M'_i[0] = 0 \ , 
M'_i[j\geq 1] = \left\{ \begin{array}{lr}
0 & \mathrm{if}\ M_i[j-1] = M_i[j] \\
1 & \mathrm{if}\ M_i[j-1] \neq M_i[j]
\end{array} \right. \ ,
\eeq
where $M_i[j]$ denotes the $j$-th measurement outcome of the $i$-th simulation.
We calculated the one-sided power spectral density (PSD) of each time series $M_i'$ followed by averaging over $i$, which for the Bernoulli process with a constant transition rate should give a flat PSD. As we show in Fig.~\ref{fig:PartiyPSD}, our simulations for the $1/f$ and quasi-static noise models exhibit a peak in their PSD, indicating that the transition rate is not constant during the measurement sequence. Such variations would then need to be taken into account in the design of error correction decoders for example (for a recent review, see Ref.~\cite{deMarti2024}). We can attribute this to the non-trivial effect of the noise on the realization of the parity measurement, where the finite-duration CNOTs and noisy ancillas results in non-trivial parity-flip statistics. Simulations where the CNOTs are instantaneous, the ancilla qubit is uncorrupted, and where we apply the identity operation for the data qubits for 720ns (the same total duration of the two finite-time CNOTs) exhibit a flatter PSD for the quasi-static noise model.  We show this in Appendix~\ref{app:noisymeasurement}.

\section{Computational efficiency} \label{sec:efficiency}
We briefly discuss aspects of the computational efficiency that are specific to our temporal coarse graining method.  Given the parameters of the stochastic processes that comprise the noise model, the ideal Hamiltonian $H(t)$, and a coarse time grid, most of the terms appearing in Eq.~\eqref{eqt:LeadingOrderCumulantBridge} can be pre-computed independently of any specific noise trajectory on the coarse-time grid. These noise-trajectory-independent pieces can be pre-computed once and stored for use in repeated realizations of the simulation. For each noise operator $B_{\alpha}(t)$ with associated stochastic process $\eta_{\alpha}(t)$ that is expressed in terms of $N_{\alpha}$ OU processes, we have $N_{\alpha}$ contributions to each of $\overline{\tilde{\mL}_1^{(\mathrm{S})}(t_k,t_{k-1})^2}$ and $\overline{\tilde{\mL}_2^{(\mathrm{S})}(t_k,t_{k-1})}$ that only depend on the two point correlation of the OU zero-boundary bridge process, which is specified by the properties of the OU process. This only scales as $N_{\alpha}$ for each noise operator even for $\overline{\tilde{\mL}_2^{(\mathrm{S})}(t_k,t_{k-1})}$ because the different OU zero-boundary bridge processes are independent. Thus the cost of this pre-computation scales as $\sum_{\alpha} N_{\alpha}$.

The term $\tilde{\mL}_1^{(\mathrm{D})}(t_k,t_{k-1})$ can be expressed as a linear function of the noise trajectory values at $t_k$ and $t_{k-1}$, and the coefficients of these functions can be expressed in terms of the properties of the OU processes.  Thus, the cost of this pre-computation scales as $\sum_{\alpha} N_{\alpha}$.

The term $ \tilde{\mL}_2^{(\mathrm{D})}(t_k,t_{k-1})$ is a quadratic function of the noise trajectory values at $t_k$ and $t_{k-1}$. The coefficients of these functions can be expressed in terms of the properties of the OU processes, but because there is no ensemble averaging in these terms we cannot take advantage of the OU processes being independent. Therefore, the cost of this pre-computation scales as $\left(\sum_{\alpha} N_{\alpha}\right)^2$. This can become costly when the total number of OU processes is large, giving a disadvantage relative to the filter function approach, for example.

As in our example in Sec.~\ref{sec:Weight2Parity}, the terms associated with circuit elements that appear repeatedly (the two CNOT operations) only need to be pre-computed \emph{once} since these terms do not depend on where the circuit element appears in the circuit.  The concatenation of circuit elements is performed straightforwardly as in Eq.~\eqref{eqt:concatenation}.

Thus, once this pre-computation is performed, the cost of a simulation only includes: (1) generating a noise realization on the coarse time grid, (2) applying Eq.~\eqref{eqt:LeadingOrderCumulant} using the decomposition in Eq.~\eqref{eqt:LeadingOrderCumulantBridge} using the generated noise realization, and (3) repeating steps (1) and (2) $N_{\mathrm{MC}}$ times to achieve a desired statistical accuracy. The Monte Carlo overhead of performing $N_{\mathrm{MC}}$ may be large, but independent noise trajectory simulations can be performed in parallel, reducing the temporal overhead of $N_{\mathrm{MC}}$ by a factor that only depends on the available computing resources.

While we have presented our approach using the superoperator formalism in Liouville space, the approach could just as well be implemented in Hilbert space. For a $d$-dimensional system, the former approach requires multiplying $d^2 \times d^2$ matrix representations of the process matrices with a $d^2$ vector representation of the density matrix. The latter approach instead requires left-right multiplication of $d \times d$ matrices with the $d \times d$ representation of the density matrix, so it may provide some efficiencies depending on the value of $d$.

\section{Conclusions} \label{sec:conclusions}
%
In this work, we have developed an approach for implementing temporally coarse-grained dynamics.  The approach relies on a Monte Carlo aspect, where realizations of the noise processes are generated over a coarse time grid, and conditioning the dynamics on a noise realization. The conditioned process between two time points can then be expressed in terms of a deterministic function and a (stochastic) zero-boundary bridge process. The zero-boundary bridge processes associated with the different (non-overlapping) pair of time points are independent, and we perform an ensemble average over the zero-boundary bridge processes.  This effectively integrates out the high-frequency components (short timescales) of the noise, while the correlations over longer timescales are mediated by the coarse noise realization.

We focus on noise processes that can be expressed as a sum of OU processes, where (1) we can express the deterministic function and the two-point correlation of the zero-boundary bridge process analytically, (2) the stochastic component of the zero-boundary bridge process is independent of coarse noise realization, and (3) we can generate noise trajectories exactly. This means that there are three sources of error in our approach: the order at which we truncate our Magnus and cumulant expansion, the numerical accuracy of the integrals in the Magnus expansion terms, and sampling errors from averaging over a finite number of noise trajectories. 

While OU processes are a convenient choice for us, we expect the general formalism to be applicable to other Markovian processes, although some simplifications may be lost. As noted, when using an OU process, the zero-boundary bridge process between time points $t_{k-1}$ and $t_k$ is independent of the two values of coarse noise trajectory at those time points, requiring computation of only a single noise propagator.  However, if we were to consider a telegraph process \cite{Kac1974}, which is Markovian but non-Gaussian, the zero-boundary bridge process as defined in Eq.~\eqref{eqt:OUBridgeProcessMain} would depend on the value of the process at $t_{k-1}$ and $t_k$.  Nevertheless,  there would only be a finite number of values at these time points to consider for such a case, so their pre-computation remains feasible. 

As noted at the end of Sec.~\ref{sec:Method}, there is a trade-off between accuracy and efficiency in the choice of the coarse-graining time step. A larger time step has the advantage of averaging more of the noise, which possibly reduces the amount of Monte Carlo averaging needed, at the cost of incurring a larger error arising solely from the truncation of the Magnus expansion. Whether the advantage is large enough to compensate for the additional cost of density matrix evolution and outperform other methods, for example pure-state simulation with a potentially significantly smaller time step, likely depends on the noise characteristics (the decomposition into OU processes and the range of the OU frequencies), the system size, and the circuit depth. We therefore expect our approach to be a useful addition to, rather than a replacement of, other open system simulation methods.

We have demonstrated our approach with a few examples.  With the simple single-qubit and two-qubit examples, we are able to explicitly show how the ensemble average over the zero-boundary bridge processes results in decoherence-type dynamics. This fits with our intuitive picture of this averaging being effectively like integrating over the high-frequency components of the noise. 
Our first non-trivial example is a simulation of the full-permutation dynamical decoupling protocol for a three-spin exchange-only qubit, where we show how our method correctly captures the effect of magnetic field gradients generated by the magnetic noise.
We further highlight the advantages of our approach, namely that concatenation is trivial without any additional cost in the presence of mid-circuit measurements, using the much more complicated example of sequential parity check measurements using three qubits in a singlet-triplet encoding (total of six spins).  In this example, we are able to simulate many measurements in order to extract the temporal fluctuations in the parity-flip rate.
While our analysis of the different features of the parity measurements allows us to discriminate between our very different noise models, it is not clear whether this measurement alone provides a useful form of noise spectroscopy for noise models that are more similar.  It would be interesting to study whether other quantum error correction-type measurements give more discriminatory power, and we leave this for future work.

\begin{acknowledgments}
We acknowledge useful technical discussions with Kevin Young.

This work was performed, in part, at the Center for Integrated Nanotechnologies, an Office of Science User Facility operated for the U.S. Department of Energy (DOE) Office of Science.

This article has been co-authored by an employee of National Technology \& Engineering Solutions of Sandia, LLC under Contract No. DE-NA0003525 with the U.S. Department of Energy (DOE). The employee owns all right, title and interest in and to the article and is solely responsible for its contents. The United States Government retains and the publisher, by accepting the article for publication, acknowledges that the United States Government retains a non-exclusive, paid-up, irrevocable, world-wide license to publish or reproduce the published form of this article or allow others to do so, for United States Government purposes. The DOE will provide public access to these results of federally sponsored research in accordance with the DOE Public Access Plan \url{https://www.energy.gov/downloads/doe-public-access-plan.}
\end{acknowledgments}

\bibliographystyle{quantum}
\bibliography{refs}

\newpage
\appendix
\section{Explicit Expressions for Second Order Cumulant Expansion Terms} \label{app:Cumulant}
For the derivation in Sec.~\ref{sec:Review}, we express the second order terms of the superoperator $\mK(t,t_0)$ (Eq.~\eqref{eqt:LeadingOrderCumulant}) in terms of the Hermitian orthonormal basis $\left\{ P_k \right\}_{k=0}^{d^2 - 1}$ with respect to the Hilbert-Schmidt inner product: $\Tr \left( P_k P_l \right) = \delta_{kl}$ basis, such that $\mK_{ij}(t,t_0) = \Tr \left( P_i \mK(t,t_0)(P_j) \right)$.  The terms are given by:
\bes \label{eqt:DeltaGamma}
\begin{align}
-i \overline{\tilde{\mL}_1(t,t_0)} &= -i \sum_{\alpha} \sum_k \int_{t_0}^t dt_1 \overline{\eta_{\alpha}(t_1)}  \nonumber \\
& \times \tilde{\mathcal{B}}_{\alpha,k}(t_1,t_0) \left[P_k, \odot\right] = 0 \ , \\
-i \overline{\tilde{\mL}_2(t,t_0)} &= - \frac{1}{2} \sum_{\alpha, \beta} \sum_{k,l} \int_{t_0}^t dt_1 \int_{t_0}^{t_1} dt_2 \ \overline{\eta_\alpha(t_1) \eta_{\beta}(t_2)} \nonumber \\
& \times \tilde{\mB}_{\alpha k}(t_1,t_0) \tilde{\mB}_{\beta l}(t_2,t_0) \left[ \left[P_k, P_l \right], \odot \right] \nonumber \\
&= -\frac{1}{2}  \sum_{\alpha, \beta} \sum_{k,l} \Delta_{\alpha \beta, kl}(t,t_0) \left[ \left[P_k, P_l \right], \odot \right]  \ , \\
-\frac{1}{2} \overline{\tilde{\mL}_1(t,t_0)^2} &= -\frac{1}{2} \sum_{\alpha \beta} \sum_{kl} \int_{t_0}^t dt_1 \int_{t_0}^t dt_2 \ \overline{\eta_\alpha(t_1) \eta_{\beta}(t_2)} \nonumber \\
& \times \tilde{\mB}_{\alpha k}(t_1,t_0) \tilde{\mB}_{\beta l}(t_2, t_0) \left[ P_k, \left[P_l,\odot \right]\right] \nonumber \\
&= -\frac{1}{2} \sum_{\alpha \beta} \sum_{kl} \Gamma_{\alpha \beta, kl}(t,t_0) \left[ P_k, \left[P_l,\odot \right]\right] \ .
\end{align}
\ees
where we have assumed that $\overline{\eta_{\alpha}(t)} = 0$. The matrix representation of these terms requires us to calculate:
\bes
\begin{align}
  \Tr \left( P_i  \left[ \left[P_k, P_l \right], P_j \right] \right) & = f_{ijkl} \ , \\
  \Tr \left( P_i \left[ P_k, \left[P_l,P_j \right]\right] \right) &=  g_{ijkl} \ .
\end{align}
\ees
Using this notation the matrix representation of $\mK(t,t_0)$ is given by:
\begin{eqnarray} \label{eqt:cumulantmatrixelements}
\mK_{ij}(t,t_0) &=& - \frac{1}{2} \sum_{\alpha,\beta} \sum_{k,l} \left( f_{ijkl} \Delta_{\alpha \beta, kl}(t,t_0) \right. \nonumber \\
&& \left. + g_{ijkl} \Gamma_{\alpha \beta, kl}(t,t_0) \right) \ ,
\end{eqnarray}
agreeing with the derivation in Ref.~\cite{Hangleiter2021}. 

In the case of the conditioned process that we discuss in Sec.~\ref{sec:BridgeEnsembleAverage}, we replace $\eta_{\alpha}(t)$ in $H_{\mathrm{N}}(t)$ by $\sum_{n} \left( \eta_{\alpha n k}^{(\mathrm{D})}(t) + \eta_{\alpha n k}^{(S)}(t) \right)$ for each time increment $t \in \left[t_k, t_{k-1} \right]$, which is the sum of the deterministic function and the stochastic zero-boundary bridge process in the time interval $[t_k, t_{k-1}]$.  We can then express the Magnus terms $\tilde{\mL}_1$ and $\tilde{\mL}_2$ in terms of their dependence on $\eta_{\alpha n k}^{(\mathrm{D})}(t)$ and $\eta_{\alpha n k}^{(\mathrm{S})}(t)$, as shown in Eq.~\eqref{eqt:MagnusDS}.  For example,
\begin{eqnarray} 
\tilde{\mL}_2^{(\mathrm{S,D})}(t_k,t_{k-1}) &=& -  \frac{i}{2} \sum_{n,n'} \sum_{\alpha,\alpha'} \sum_{\ell,\ell'=1}^{d^2}  \left( \int_{t_{k-1}}^{t_k} d\tau_1   \right. \nonumber \\
&& \hspace{-2cm} \left. \eta_{\alpha n k}^{(\mathrm{S})}(\tau_1)  \tilde{\mB}_{\alpha \ell}(\tau_1,t_{k-1}) \int_{t_{k-1}}^{\tau_1} d\tau_2 \eta_{\alpha' n' k}^{(\mathrm{D})}(\tau_2) \right. \nonumber \\
&& \hspace{-2cm} \left. \tilde{\mB}_{\alpha' \ell'}(\tau_2,t_{k-1}) \right)  \left[ \left[ P_\ell , P_{\ell'} \right], \odot \right] \ .
\end{eqnarray}
Because only the terms $\eta_{\alpha n k}^{(S)}(t)$ are stochastic, the ensemble average over the bridge processes only affects these terms.  We can therefore write the bridge ensemble average of the terms in Eq.~\eqref{eqt:MagnusDS} as:
\bes \label{eqt:A4}
\begin{align}
     \overline{\tilde{\mL}_1(t_k,t_{k-1})} &= {\tilde{\mL}^{(\mathrm{D})}_1(t_k,t_{k-1})}  \\
     \overline{\tilde{\mL}_2(t_k,t_{k-1})} &=  {\tilde{\mL}^{(\mathrm{D})}_2(t_k,t_{k-1})} +\overline{\tilde{\mL}^{(\mathrm{S})}_2(t_k,t_{k-1})} \ , \\
    \overline{\tilde{\mL}_1(t_k,t_{k-1})^2} &=  \tilde{\mL}^{(\mathrm{D})}_1(t_k,t_{k-1})^2 + \overline{\tilde{\mL}^{(\mathrm{S})}_1(t_k,t_{k-1})^2} \ . \label{eqt:A4c}
\end{align}
\ees
In contrast to Eq.~\eqref{eqt:DeltaGamma}, the conditioned case includes a contribution from $\eta^{(\mathrm{D})}_{\alpha n k(t)}$ at first order. There are no terms that mix the two types of terms at second order because $\overline{\eta^{(\mathrm{S})}_{\alpha n k} }\eta^{(\mathrm{D})}_{\alpha' n' k'} = 0$. 

Because the expressions in Eq.~\eqref{eqt:A4} have non-zero mean generically, we must use the cumulant average when calculating the cumulant expansion (Eq.~\eqref{eqt:CumulantAverage}). For example, we have:
\begin{eqnarray}
 \overline{\tilde{\mL}_1(t_k,t_{k-1})^2}^c &=&  \overline{\tilde{\mL}_1(t_k,t_{k-1})^2} - \overline{\tilde{\mL}_1(t_k,t_{k-1})}^2 \nonumber \\
 &=&   \overline{\tilde{\mL}^{(\mathrm{S})}_1(t_k,t_{k-1})^2} \ .
\end{eqnarray}
Thus, using the cumulant average, the $\tilde{\mL}^{(\mathrm{D})}_1(t_k,t_{k-1})^2$ term in Eq.~\eqref{eqt:A4c} is eliminated, and the cumulant expansion using the bridge ensemble average then takes the form given in Eq.~\eqref{eqt:LeadingOrderCumulantBridge}.
\section{Background on Ornstein-Uhlenbeck Processes} \label{app:OU}
An OU process $X_t$ is a stochastic process satisfying the stochastic differential equation \cite{Uhlenbeck1930}
\beq \label{eqt:SDE}
d X_t= \gamma \left( \mu - X_t  \right) dt+ \sigma dW_t \ ,
\eeq
where $\gamma, \sigma > 0$ and $\mu$ are parameters characterizing the process and $W_t$ denotes a Wiener process. To solve this equation analytically, one goes through the following steps.  Let us define the function $f(X_t, t) = X_t e^{\gamma t}$.  From this we get:
\begin{eqnarray}
d f(X_t, t) &=& \gamma X_t e^{\gamma t} dt + e^{\gamma t} d X_t \nonumber \\
&=& \gamma X_t e^{\gamma t} dt + e^{\gamma t} \left( \gamma (\mu - X_t ) dt + \sigma dW_t \right) \nonumber \\
&=& \mu \gamma e^{\gamma t} dt + e^{\gamma t} \sigma d W_t \ .
\end{eqnarray}
Conditioned on a value of $x_0$ at time $t = 0$, we can integrate this expression from $0$ to $t$ to give:
\beq
X_t e^{\gamma t}  - x_0 = \mu \gamma \int_0^t ds  e^{\gamma s}  + \sigma \int_0^t e^{\gamma s} \sigma d W_s \ .
\eeq
Rewriting, we get:
\beq
X_t = x_0 e^{-\gamma t} + \mu \left( 1 - e^{-\gamma t} \right) + \sigma e^{-\gamma t} \int_0^t e^{\gamma s} d W_s \ .
\eeq 
We now use the fact that 
\beq
\int_0^t f(s) d W_s \sim \mathcal{N} \left( 0 , \int_0^t f(s)^2 \right) \ ,
\eeq
which for us gives:
\beq
\int_0^t e^{\gamma s} d W_s \sim \mathcal{N} \left( 0 , \frac{1}{2 \gamma} \left(e^{2 \gamma t} - 1 \right) \right) \ ,
\eeq
which has the same distribution as $\frac{1}{\sqrt{2 \gamma}} W_{e^{2 \gamma t} - 1}$.  Therefore, the analytical solution for the process is given by:
\begin{eqnarray} \label{eqt:OUformalsol}
\left( X_t | X_0 = x_0 \right) &=&  x_0 e^{-\gamma t} + \mu \left( 1 - e^{-\gamma t} \right) \nonumber \\
&& \hspace{-1cm} + \frac{\sigma}{\sqrt{2 \gamma}} e^{-\gamma t} W_{e^{2 \gamma t} -1} \ , \quad t \geq 0 \ .
\end{eqnarray}
Conditioned on $x_0$, this gives for $t, s \geq 0$:
\bes \label{eqt:ConditionedOUExp}
\begin{align}
\mathbb{E} \left[ X_t | X_0 = x_0 \right] &= x_0 e^{-\gamma t} + \mu \left( 1 - e^{- \gamma t} \right) \ ,  \label{eqt:mean} \\
\mathrm{Cov} \left( X_t, X_s | X_0 = x_0 \right) &= \frac{\sigma^2}{2 \gamma} \left( e^{-\gamma |t - s|} - e^{-\gamma(t+s)} \right) \ , \\
\mathrm{Var} \left[ X_t | X_0 = x_0 \right] &= \frac{\sigma^2}{2 \gamma} \left( 1 - e^{-2 \gamma t} \right) \ . \label{eqt:var}
\end{align}
\ees
If we now remove the condition on the initial state and take $X_0 \sim \mathcal{N}(\mu, \sigma^2/(2 \gamma))$, then the unconditioned expectation values for $t,s \geq 0$ are given by:
\bes
\begin{align}
\mathbb{E} \left[ X_t \right] &= \mu \ , \\
\mathrm{Cov} \left( X_t, X_s  \right) &= \frac{\sigma^2}{2 \gamma} e^{-\gamma |t - s|} \ ,
\end{align}
\ees
where to calculate the last expression, we use that $\mathbb{E} \left[ W_t  W_s \right] = \min(t,s)$. Since the process is wide-sense stationary, the two-sided power spectral density (PSD) is then given by the Fourier transform of the auto-covariance function defined as $\mathrm{Cov} \left( X_t, X_s  \right)$ by the Wiener–Khinchin theorem~\cite{Wiener1930,Khintchine1934}.  We have
\begin{eqnarray} \label{eqt:doublesidedPSD}
S_2(f) &=& \int_{-\infty}^{\infty} dt e^{-i 2 \pi f t} \mathrm{Cov} \left( X_t, X_s  \right) \nonumber \\
&=& \frac{\pi \sigma^2}{ \gamma} \left( \frac{1}{\pi} \frac{ \gamma}{\gamma^2 + ( 2 \pi f)^2} \right)  \ .
\end{eqnarray}
Notice we have defined this in terms of $f$ and not $\omega = 2 \pi f$, such that $\mathrm{Cov} \left( X_t, X_s  \right) = \int_{-\infty}^{\infty} df S_2(f) e^{i 2 \pi f t}$. Because we have the same power for both positive and negative arguments, the single-sided PSD is two times the two-sided PSD except at $f=0$
\begin{eqnarray}
S(f) &=& \frac{2 \pi \sigma^2}{ \gamma} \left( \frac{1}{\pi} \frac{ \gamma}{\gamma^2 + ( 2 \pi f)^2} \right) \nonumber \\
&=&  \frac{1}{2 \pi^2} \frac{\sigma^2}{ f_0^2 + f^2} \ , \quad 0 < f < \infty \ ,
\end{eqnarray}
where we defined $\gamma = 2 \pi f_0$.  

The process is Gaussian, so at any time $t$, it is described by a Gaussian random variable with mean given by Eq.~\eqref{eqt:mean} and variance given by Eq.~\eqref{eqt:var}.  This allows us to define an exact numerical time step~\cite{Gillespie1996}, as given in Eq.~\eqref{eqt:OUalgorithmMain}. 

\begin{figure}[!t] 
   \centering
   \includegraphics[width=3.25in]{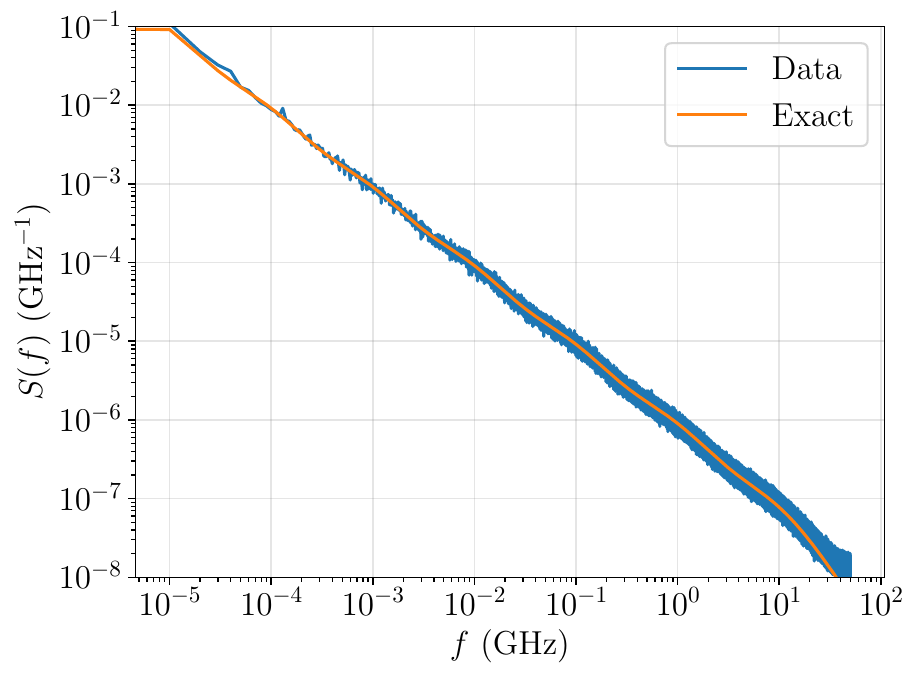} 
   \caption{PSD of a dimensionless stochastic process composed of a sum of 14 OU processes distributed uniformly on a log scale between $f=1$mHz and $f=10$GHz with $\sigma_k^2 = 2 \pi f_k  \cdot 4 \times 10^{-6} $ (these parameters correspond to the noise parameters for the charge noise in the main text; notice that here $\sigma_k^2$ has units of Hz). The simulations use Eq.~\eqref{eqt:OUalgorithmMain} with $\Delta t = 10^{-2}$ns to generate the noise realizations with a maximum simulation time of $t_f = 0.1$ms.  The PSD shown is the average over 100 independent noise trajectories.  The maximum time of the simulation explains the roll-off we observe at a frequency of $10^{4}$Hz.  The exact PSD is given by Eq.~\eqref{eqt:1fasOUspectrum}.}
   \label{fig:PSD}
\end{figure}

In order to generate a $1/f$ spectrum from a sum of OU processes, we choose our $N$ $f_k$'s to be spaced uniformly on a log scale between $f_{\min}$ and $f_{\max}$, and we then choose $\sigma^2(\gamma_k) = p \gamma_k = p 2 \pi f_k$ such that:
\beq \label{eqt:1fasOUspectrum}
S(f) = \sum_k \frac{1}{\pi} \frac{ p f_k}{f_k^2 + f^2} \ .
\eeq
To demonstrate that our noise generation is correct, we show the PSD of trajectories using Eq.~\eqref{eqt:OUalgorithmMain} with these choices in Fig.~\ref{fig:PSD}.

\section{Background on Ornstein-Uhlenbeck Bridge Processes} \label{app:OUBridge}
An OU process $X_t$ conditioned on specific boundary values, denoted $\left( X_t  | X_{t_{k-1}} = x_{k-1} , X_{t_k} = x_k \right)$ for $t_{k-1} \leq t \leq t_k$ is called an OU bridge process.  Let $\Delta t = t_k - t_{k-1}$. By fixing the boundary values of $X_t$, we have specific values for the Wiener process conditioned on the values of the OU process in Eq.~\eqref{eqt:OUformalsol}:
\begin{eqnarray}
W_{e^{2 \gamma \Delta t} - 1} | \left( X_{t_k} = x_k, X_{k-1} = x_{k-1} \right) \nonumber \\
&&  \hspace{-6.25cm} = \frac{\sqrt{ 2 \gamma}}{\sigma}  \left[ e^{\gamma \Delta t}  x_k - x_{k-1}  - \mu \left( e^{\gamma \Delta t} - 1 \right)  \right] \equiv w_k \ .
\end{eqnarray}
The conditioned Wiener process $W_{e^{2 \gamma \Delta t} - 1}$ is an OU bridge process.  We can model the conditioned Wiener process as follows (let $u_t = e^{2 \gamma (t - t_{k-1})}-1 \in [0 , u_{t_k}]$):
\beq \label{eqt:OUBridge}
B_{u_t} = \frac{u_t}{u_{t_k}} w_k + \tilde{W}_{u_t} - \frac{u_t}{u_{t_k}} \tilde{W}_{u_{t_k}} \ ,
\eeq
where the Wiener process $\tilde{W}$ is an independent Wiener process. This expression satisfies $B_{0} = \tilde{W}_{0} = 0$ and $B_{u_{t_k}} = w_k$ as desired.  Therefore, the conditioned OU process can be entirely described for $ t \in [t_{k-1},t_k ] $:
\begin{eqnarray} \label{eqt:ConditionedOU}
\left( X_t  | X_{t_{k-1}} = x_{k-1} , X_{t_k} = x_k \right) \nonumber \\
&& \hspace{-4cm} = x_{k-1} e^{-\gamma (t-t_{k-1})} + \mu \left( 1 - e^{-\gamma (t-t_{k-1})} \right) \nonumber \\
&& \hspace{-4cm}  + \frac{\sigma}{\sqrt{2 \gamma}} e^{-\gamma (t-t_{k-1})} B_{u_t} \ .
\end{eqnarray}
For simplicity we focus on the case of $\mu = 0$. It then follows that for $ t \in \left[t_{k-1}, t_k \right] $:
\begin{eqnarray} \label{eqt:EConditionedOU}
\eta^{(\mathrm{D})}_k(t) \equiv \mathbb{E} \left[ X_t | X_{t_{k-1}} = x_{k-1}, X_{t_k} = x_k \right] \nonumber \\
&& \hspace{-6cm} = x_{k-1} e^{-\gamma (t-t_{k-1})} + e^{-\gamma(t-t_{k-1})} \frac{\sigma}{\sqrt{2 \gamma}} \mathbb{E} \left[ B_{u_t} \right] \nonumber \\
&& \hspace{-6cm} = x_{k-1} e^{-\gamma (t-t_{k-1})} + e^{-\gamma(t-t_{k-1})} \frac{\sigma}{\sqrt{2 \gamma}}  \frac{u_t}{u_{t_k}} w_k \nonumber \\ 
&& \hspace{-6cm} =  x_{k-1} e^{-\gamma (t-t_{k-1})} + \frac{e^{2 \gamma(t-t_{k-1})}-1}{e^{2 \gamma \Delta t}-1} \left( e^{\gamma (t_k- t)} x_k \right. \nonumber \\
&& \hspace{-6cm} \left. -  e^{-\gamma(t-t_{k-1})} x_{k-1} \right) \ .
\end{eqnarray}
This expectation value depends on the noise realization $\lbrace x_{k-1}, x_{k} \rbrace$ and hence encodes the temporal correlations of the given realization of the OU process.

We can define a process $\left( X'_t | X_{t_{k-1}=x_{k-1}}, X_{t_k} = x_k\right)$ over the interval $ t \in \left[t_{k-1}, t_k \right] $
\begin{eqnarray}
\left( X'_t | X_{t_{k-1}}=x_{k-1}, X_{t_k} = x_k\right)  \nonumber \\
&& \hspace{-4.25cm} = \left( X_t | X_{t_{k-1}} = x_{k-1}, X_{t_k} = x_k\right) \nonumber \\
&& \hspace{-4.25cm} -  \mathbb{E} \left[ X_t | X_{t_{k-1}} = x_{k-1}, X_{t_{k}} = x_k \right]\nonumber \\
&& \hspace{-4.25cm} = \frac{\sigma}{\sqrt{2 \gamma}} e^{- \gamma(t - t_{k-1})} \left( \tilde{W}_{u_t} - \frac{u_t}{u_{t_k}} \tilde{W}_{u_{t_k}} \right)  \ .
\end{eqnarray}
The process $\left( X'_t | X_{t_{k-1}=x_{k-1}}, X_{t_k} = x_k\right)$ also describes a bridge process, but now where the process is zero at both ends.  Furthermore, it does not depend on the specific realization $\lbrace x_{k-1}, x_{k} \rbrace$ of the OU process.

If we denote $\eta_k^{(\mathrm{S})}(t) \equiv \left( X'_t | X_{t_{k-1}=x_{k-1}}, X_{t_k} = x_k\right)$, then we have
\bes  \label{eqt:BridgeOU}
\begin{align}
\overline{\eta_k^{(\mathrm{S})}(t) }  & = 0 \ , \\
\overline{\eta_k^{(\mathrm{S})}(t) \eta_k^{(\mathrm{S})}(s) } &= \frac{\sigma^2}{\gamma} \frac{ \sinh(\gamma (s - t_{k-1})) \sinh( \gamma (t_k-t))}{\sinh(\gamma \Delta t)} \ , \nonumber \\
& \hspace{1cm} t_{k-1} \leq s \leq t \leq t_k \ , \\
& \hspace{-2cm} \overline{\eta_k^{(\mathrm{S})}(t)  \eta_k^{(\mathrm{S})}(s)  \eta_k^{(\mathrm{S})}(u) }  = 0 \ ,
\end{align}
\ees
where we use $\overline{\tilde{W}_{t} \tilde{W}_{s}} = \min(t,s)$.  Notice from the covariance result that
$\eta_k^{(\mathrm{S})}(t)$ is \emph{not} wide-sense stationary, in the sense that it is invariant under shifting $s$ and $t$ by the same amount.  The covariance is invariant only under shifting all variables $s,t,t_{k},t_{k-1}$ by the same amount.
%
\section{Two-Qubit Operator Basis} \label{app:2qOperatorBasis} 

We show how we can recover Eq.~\eqref{eqt:2QubitDephasing} using the formalism presented in Sec.~\ref{sec:Method}. We define the eigenbasis of $S^z$ to be given by the spin-up $\ket{\!\!\uparrow}$ and spin-down $\ket{\!\!\downarrow}$ states such that $S^z \ket{\! \uparrow} = \frac{1}{2} \ket{\! \uparrow}$ and $S^z \ket{\! \downarrow}= -\frac{1}{2} \ket{\! \downarrow}$. We choose to work in the eigenbasis of the Hamiltonian given by the singlet and triplet states:
\bes
\begin{align}
\ket{S_0} &= \frac{1}{\sqrt{2}} \left( \ket{\! \uparrow \downarrow} - \ket{\! \downarrow \uparrow} \right) \ ,  \\
\ket{T_0} &= \frac{1}{\sqrt{2}} \left( \ket{\! \uparrow \downarrow} + \ket{\! \downarrow \uparrow} \right)  \ , \\
\ket{T_-} &= \ket{\! \downarrow \downarrow} \ , \\
 \ket{T_+} &= \ket{\! \uparrow \uparrow} \ .
\end{align}
\ees
 We write the ideal evolution operator in the interval $t \in [t_k, t_{k-1}]$ as:
\begin{eqnarray}
U_I(t,t_{k-1}) &=& e^{i \frac{3}{4} J (\tau-t_{k-1})} \ketbra{S_0}{S_0} \oplus e^{-i\frac{1}{4} J(\tau-t_{k-1})} \nonumber \\
&& \hspace{-0.5cm} \times  \left( \ketbra{T_-}{T_-}  +  \ketbra{T_0}{T_0} + \ketbra{T_p}{T_p} \right) \ ,
\end{eqnarray}
which suggests that a convenient choice of operator basis is one where we have:
\bes
\begin{align}
P_0 &= \ketbra{S_0}{S_0} \ , \\
P_1 &= \frac{1}{\sqrt{3}} \left( \ketbra{T_-}{T_-}  + \ketbra{T_0}{T_0} + \ketbra{T_+}{T_+} \right) \ ,
\end{align}
\ees 
such that the only non-zero $\tilde{\mathcal{B}}_k$ terms are given by:
\bes
\begin{align}
\tilde{\mathcal{B}}_0(\tau, t_0) & = -\frac{3}{4} J \ , \\
\tilde{\mathcal{B}}_1(\tau, t_0) & = \frac{\sqrt{3}}{4} J \ .
\end{align}
\ees
The remaining operators can be built using the Gell-Mann basis and terms that map between the two subspaces. The choice uses the Gell-Mann basis in the $S=1$ three-dimensional subspace, and the rest of the terms in the basis correspond to operators that map between the $S=0$ and $S=1$ subspaces.  The operators are given by:  
\bes
\begin{align}
P_0 &= \ketbra{S_0}{S_0} \ , \\
P_1 &= \frac{1}{\sqrt{3}} \left( \ketbra{T_-}{T_-}  + \ketbra{T_0}{T_0} + \ketbra{T_+}{T_+} \right) \ , \\
P_2 &= \frac{1}{\sqrt{2}} \left( \ketbra{T_-}{T_0}  + \ketbra{T_0}{T_-} \right) \ , \\
P_3 &= \frac{1}{\sqrt{2}} \left( \ketbra{T_-}{T_+}  + \ketbra{T_+}{T_-} \right) \ , \\
P_4 &= \frac{1}{\sqrt{2}} \left( \ketbra{T_0}{T_+}  + \ketbra{T_+}{T_0} \right) \ , \\
P_5 &= \frac{1}{i\sqrt{2}} \left( \ketbra{T_-}{T_0}  - \ketbra{T_0}{T_-} \right) \ , \\
P_6 &= \frac{1}{i\sqrt{2}} \left( \ketbra{T_-}{T_+}  - \ketbra{T_+}{T_-} \right) \ , \\
P_7 &= \frac{1}{i\sqrt{2}} \left( \ketbra{T_0}{T_+} - \ketbra{T_+}{T_0} \right) \ , \\
P_8 & = \frac{1}{\sqrt{2}} \left( \ketbra{T_-}{T_-} - \ketbra{T_0}{T_0} \right) \ , \\
P_9 & = \frac{1}{\sqrt{6}} \left( \ketbra{T_-}{T_-} + \ketbra{T_0}{T_0} - 2 \ketbra{T_+}{T_+} \right) \ , \\
P_{10} & = \frac{1}{\sqrt{2}} \left( \ketbra{S}{T_-} + \ketbra{T_-}{S} \right) \ , \\
P_{11} & = \frac{1}{\sqrt{2}} \left( \ketbra{S}{T_0} + \ketbra{T_0}{S} \right) \ , \\
P_{12} & = \frac{1}{\sqrt{2}} \left( \ketbra{S}{T_+} + \ketbra{T_+}{S} \right) \ , \\
P_{13} & = \frac{1}{i\sqrt{2}} \left( \ketbra{S}{T_-} - \ketbra{T_-}{S} \right) \ , \\
P_{14} & = \frac{1}{i\sqrt{2}} \left( \ketbra{S}{T_0} - \ketbra{T_0}{S} \right)\ , \\
P_{15} & = \frac{1}{i\sqrt{2}} \left( \ketbra{S}{T_+} - \ketbra{T_+}{S} \right) \ .
\end{align}
\ees 
If we consider the case of a single OU process and condition it on a specific realization at the time points $\left\{t_k \right\}$, we find the integration over the zero-boundary bridge process gives:
\begin{eqnarray}
 \overline{\tilde{\mL}^{(\mathrm{S})}_2(t_k,t_{k-1})} &=& 0  \ ,\\
-\frac{1}{2} \overline{\tilde{\mL}^{(\mathrm{S})}_1(t_k,t_{k-1})^2} 
&=&\iint_{t_{k-1}}^{t_k} dt_1 dt_2 \ \overline{\eta^{(\mathrm{S})}(t_1) \eta^{(\mathrm{S})}(t_2)}  \nonumber \\
&& \hspace{-2cm}\times   \left( L \odot L^\dagger - \frac{1}{2} \left\{ L^\dagger L, \odot \right\} \right) \ ,
\end{eqnarray}
with $L = \tilde{\mathcal{B}}_0 P_0 +  \tilde{\mathcal{B}}_1 P_1 = J \vec{S}_1 \cdot \vec{S}_2$.  

\section{Singlet-triplet qubit parameters} \label{app:SingletTriplet}
For the three-qubit problem (6 spins) in Fig.~\ref{fig:weight2parity}, we label the spins from top to bottom as spins 1 through 6, and we assume the Hamiltonian is given as in Eq.~\eqref{eqt:6SpinH}.
We define $\frac{1}{6} \sum_{i=1}^6 g_i B_i^z = g B^z$ and pick the values $g = 2$, $B^z = 500.05$ mT corresponding to an applied $500$ mT field and a $50 \ \mu $T background magnetic field.
Denoting the relative magnetic field differences between the spins by $\Delta_{i,i+1} = \mu_{\mathrm{B}} \left( g_i B_i^z - g_{i+1} B_{i+1}^z \right)/{h}$, we choose:
\beq
\Delta_{12} = \Delta_{23} = - \Delta_{34} = - \Delta_{45} = \Delta_{56} = 10 \mathrm{MHz}  \ . 
\eeq

\begin{figure*}[!htbp] 
   \centering
   \subfigure[CNOT12]{\includegraphics[width=2.25in]{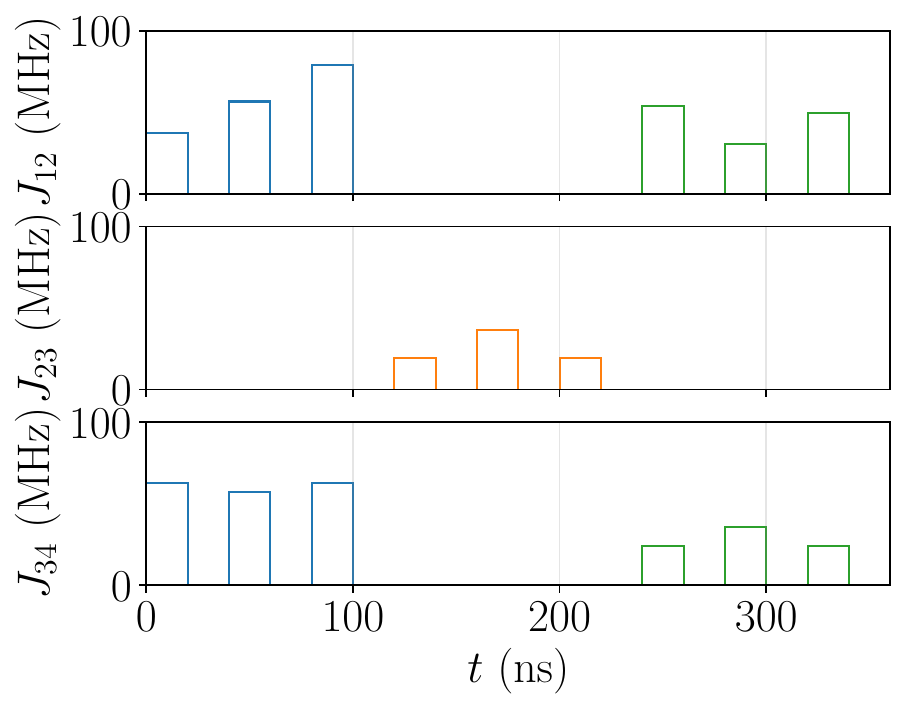} }
   \subfigure[CNOT23]{\includegraphics[width=2.25in]{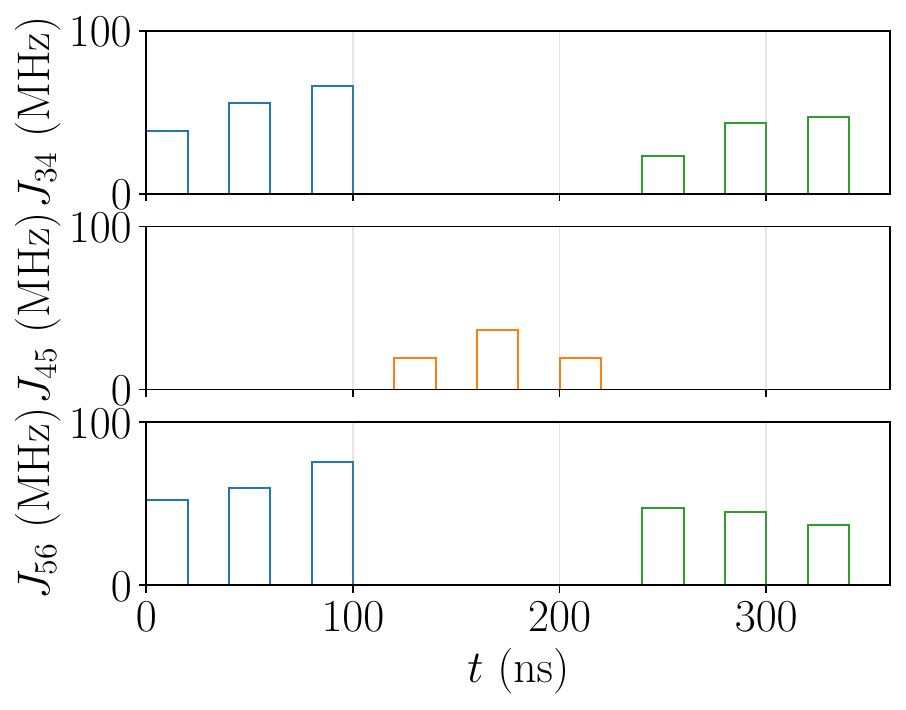}}
   \subfigure[$120$ns Identity gate]{\includegraphics[width=2.25in]{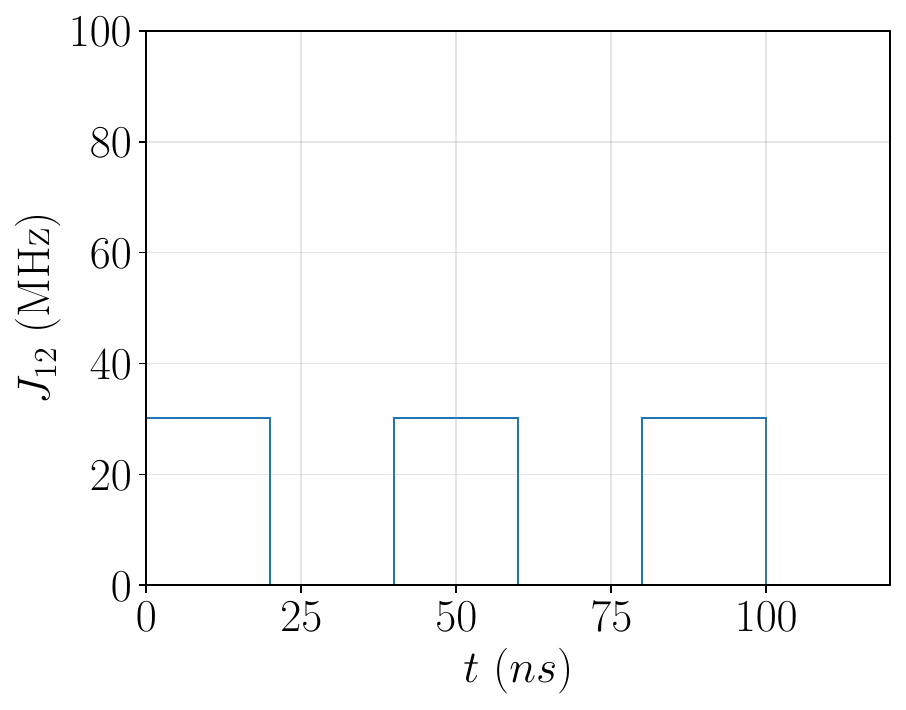}}
   \caption{CNOT and idle gate compilations. (a,b) Implementation of singlet-triplet encoded CNOT gates using finite-width exchange pulses with control$\to$target qubits 1$\to$2 and 2$\to$3, respectively. (c) Identity gate for the first encoded singlet-triplet qubit in the three-qubit array.}
   \label{fig:GateImplementation}
\end{figure*}

We show in Fig.~\ref{fig:GateImplementation} our implementation of these gates, where we fix the duration of exchange interactions to $20$ns and wait $20$ns after each exchange. For simplicity, we assume a square pulse shape and do not include any control artifacts. The CNOT gates are identified in two parts.  We first identify a pulse sequence on $J_{23}$ and $J_{45}$ that give rise to quantum gates with the same Makhlin invariant~\cite{Makhlin2002} as the CNOT gate, which then means that our quantum gate is a CNOT gate up to single qubit rotations.  We then identify pulses on $J_{12},J_{34},J_{56}$ that implement the necessary one qubit rotations to make the entire sequence a CNOT gate.  We find that three pulses is sufficient to identify the necessary gates for the parameters we have picked above. The identification of each set of pulse sequences is done using the Nelder-Mead minimization algorithm~\cite{Nelder1965} as implemented in SciPy 1.14.1 \cite{SciPy}.

\begin{figure}[!htbp] 
   \centering
   \subfigure[]{\includegraphics[width=3.25in]{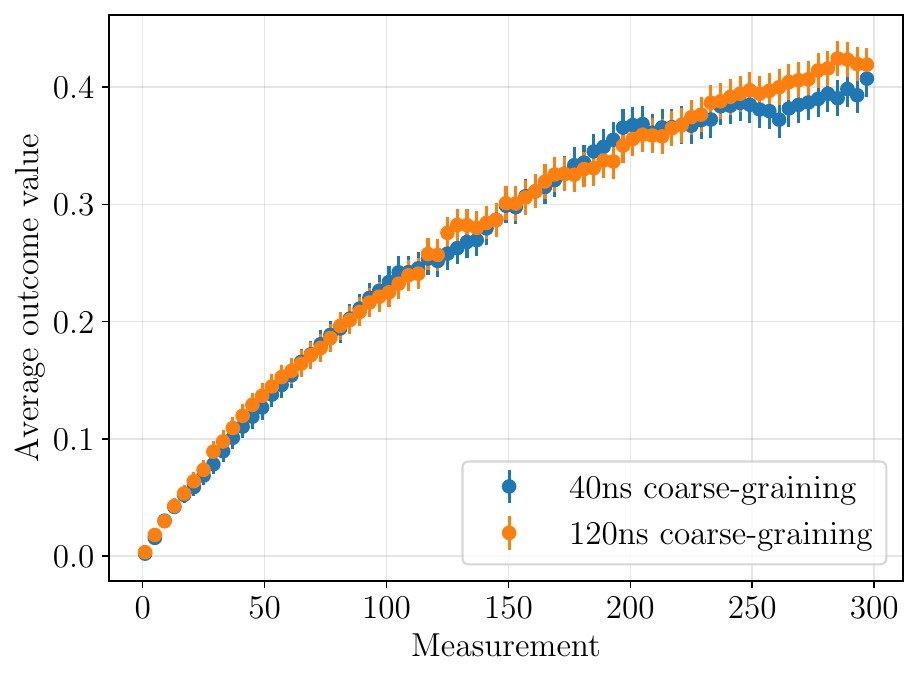}}
   \subfigure[]{\includegraphics[width=3.25in]{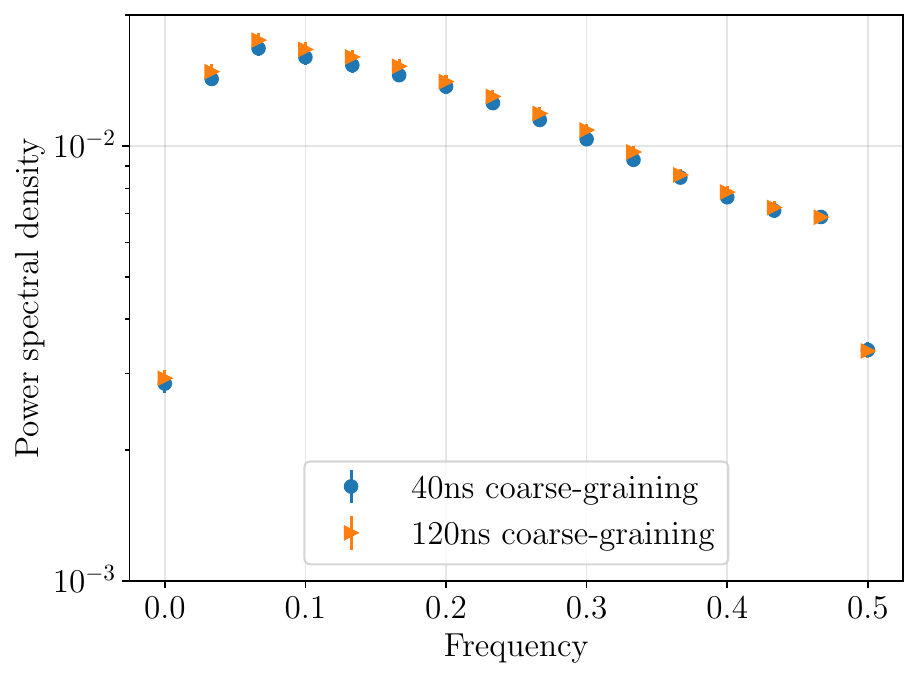}}
   \caption{Comparing the results for the $1/f$ noise model  with a $40$ns and a $120$ns coarse-graining time scale for (a) the expectation value of the measurement outcomes and (b) the PSD of the parity-flip time series for the $1/f$ noise model. The simulations use 300 measurements, and the PSD is estimated using Welch's method \cite{Welch1967} as implemented in SciPy 1.14.1 \cite{SciPy} using a segment length of 30.  The results are averaged over a total of $4 \times 10^3$ independent simulations, with the error bars  (mostly hidden by the markers) being the $2 \sigma$ confidence interval as inferred by performing a bootstrap over each simulation's estimated PSD.}
   \label{fig:CompareCoarse}
\end{figure}

Our implementation of the various gates has the feature that it gives two obvious coarse-graining time scales that allow us to only consider two spins at a time.  The first is to coarse grain over a single pulse and idle group, for a total coarse graining time scale of 40ns.  The second is to coarse grain over three sequential pulse and idle groups, for a coarse graining time scale of 120ns.  As we show in Fig.~\ref{fig:CompareCoarse} for the expectation value of measurement outcomes and the PSD of the parity flip time series, we find no statistical difference between the two, but the longer coarse graining simulations require $1/3$ less time to perform.
\section{Quasi-static Noise as a Limit of Ornstein-Uhlenbeck Processes} \label{app:OtherNoise}
We consider other stochastic processes that can be treated as limits of the OU process.  Given an OU process with parameters $\sigma$ and $\gamma$ (see Eq.~\eqref{eqt:SDE}), we take $\sigma^2 = p \gamma$ followed by the limit $\gamma \to 0$ for the OU process. We then get a process that is constant in time but whose initial value is random with $X_0 \sim \mathcal{N}(0, p/2)$.  This gives rise to our quasi-static noise model.

\section{Noise parameter tuning} \label{app:NoiseParameterTuning} 
%
For the noise models discussed in Secs~\ref{sec:EXonly} and ~\ref{sec:Weight2Parity}, we choose the noise strength parameter $p$ (Eq.~\eqref{eqt:1fasOUspectrum}) to fit the $T_2^\ast$ measurements of Ref.~\cite{Weinstein2023}. We choose the parameters of the magnetic noise $\delta \vec{B}_{k}$ to get a fixed free induction decay $T_2^\ast$ of $3.5\mu$s. The free induction decay measurement is performed on two spins and proceeds as follows.  The system is prepared in the singlet state $\ket{S_0} = \frac{1}{\sqrt{2}} \left( \ket{\! \uparrow \downarrow} - \ket{\! \downarrow \uparrow} \right)$ and allowed to evolve with the Hamiltonian $H_{\mathrm{I}} + H_{\mathrm{N}}$ with all the exchange interactions turned off and $B_0 = 50 \ \mu \mathrm{T}$. The probability of measuring the singlet state is fit to a function of the form:
\beq
\mathrm{Pr}(t) = \frac{1}{2} \left( 1 + \exp \left[ - \left(\frac{T}{T_2^\ast} \right)^b \right] \right) \ . 
\eeq

We can make analytical progress by assuming the noise Hamiltonian is given by $H_{\mathrm{N}} =  \sum_{k=1}^2 \left(g  \mu_B B_0 + \delta B_k^z \right) S_k^z $. Ignoring the fluctuations in the $x,y$ directions is a reasonable approximation with the large value of $B_0$ considered. This can be seen by observing that, in this regime of $\vert \delta B_{x,y} \vert \ll B_{0}$, the transverse magnetic noise components $x,y$ influence the Zeeman splitting of a spin only perturbatively as $\vert \delta B_{x,y} \vert^{2}/B_{0}$, while longitudinal fluctuations $\delta B_{z}$ enter at first order. The probability of observing a singlet is then given by
\beq
\mathrm{Pr}(t) = \frac{1}{2} \left( 1 + \cos \left( \frac{1}{\hbar} \int_0^t d \tau \left(\delta B_1^z(\tau) - \delta B_2^z(\tau) \right) \right) \right)  \ .
\eeq
If we assume that $\delta B_1^z$ and $\delta B_2^z$ are independent but identical stochastic processes, we can perform an ensemble average to get
\beq
\mathrm{Pr}(t) = \frac{1}{2} \left( 1 + e^{-2K(t) } \right) \ ,
\eeq
where
\beq \label{eqt:KT}
K(t) = \frac{1}{2 \hbar^2} \int_0^t dt_1 \int_0^t dt_2 \overline{ \delta B^z(t_1) \delta B^z(t_2)} \ .
\eeq
We can now relate $T_2^\ast$ to the parameters of our different noise models.
\begin{enumerate}
\item For $\delta B^z$ being a sum of $n$ OU processes with $\sigma_i^2 = p \gamma_i , i = 1, \dots, n$, then
\beq
K(t) = \frac{t}{\hbar^2} \sum_{i=1}^n \left[ \frac{ p}{2\gamma_i} \left( 1 + \frac{e^{-\gamma_i t}-1}{\gamma_i t}  \right) \right] \ .
\eeq
In the limit of $\gamma_i t \ll 1$, we can approximate $T_2^\ast \approx \sqrt{2 \hbar^2/(np)}$ with $b = 2$. For the magnetic noise processes $\left\{ \delta \vec{B}_k \right\}_{k=1}^3$, which have units of Hz when divided by $h$, each is given by a sum of 9 independent OU processes with frequencies $\left\{f_k \right\}_{k=1}^9$ distributed log-uniformly from $1$mHz to $100$kHz, with $p/h^2 = \left(  2.2 \times 10^{-5} \mathrm{GHz}\right)^2$ (see Eq.~\eqref{eqt:1fasOUspectrum}).  

\item For $\delta B^z$ being quasistatic noise, then
\beq
K(t) = \frac{2 p }{\hbar^2} t^2 \ .
\eeq
This gives $T_2^\ast =  \sqrt{2 \hbar^2 /p}$ with $b = 2$. We take $p/h^2 = (   6.431 \times 10^{-5} \mathrm{GHz})^2$.
\end{enumerate}

To characterize the noise on the exchange coupling, we consider a 3-spin system. We assume the first two spins are prepared in the singlet state $\ket{S_0}$ and without loss of generality the third spin is prepared in $\ket{\! \uparrow}$. The exchange interaction is turned turn on to $J_0$ between spins 2 and 3, and we measure the probability of measuring the singlet state for the first two spins.  The probability is fit to a function of the form:
\beq
\mathrm{Pr}(t) = \frac{5}{8} + \frac{3}{8} \cos( J_0 t / \hbar)  \exp \left[ - \left(\frac{T}{T_2^\ast} \right)^b \right] \ .
\eeq

\begin{figure}[!htbp] 
   \centering
   \subfigure[$J_{23}/h = 0$MHz]{\includegraphics[width=3.25in]{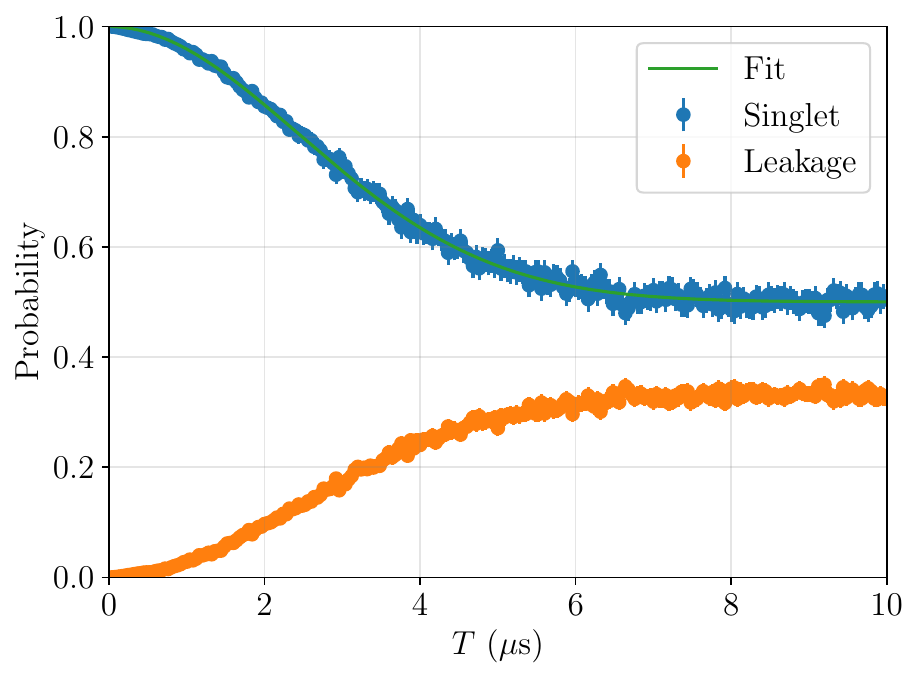} \label{fig:T2StarJ=0}}
   \subfigure[$J_{23}/h = 100$MHz]{\includegraphics[width=3.25in]{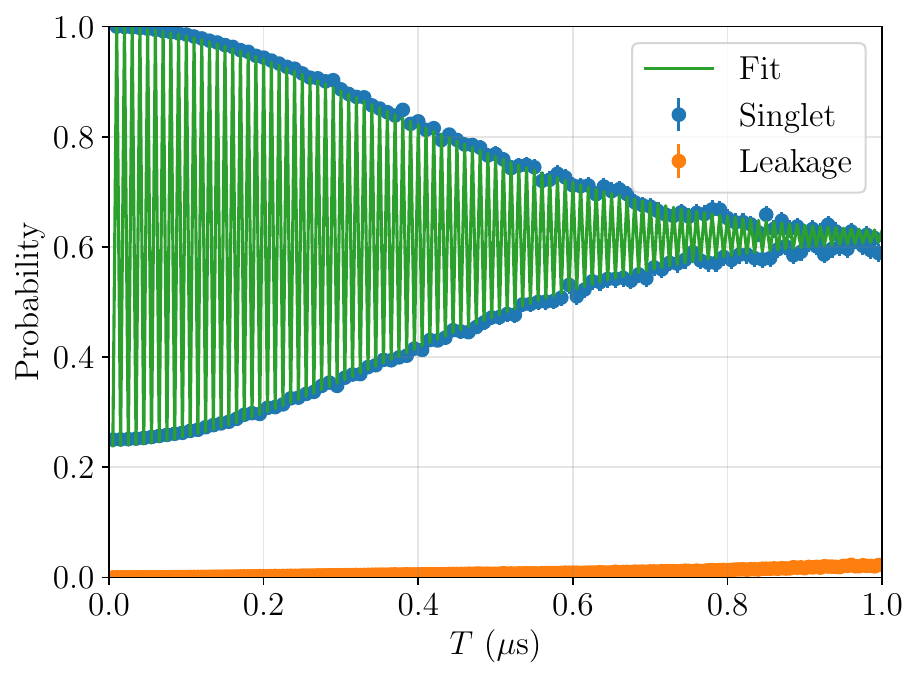}\label{fig:T2StarJ=100}}
   \caption{Free induction decay results for a 3-spin DFS qubit. (a) Probability of measuring a singlet after a time $t_{\mathrm{evol}}$ when the exchange couplings are off ($J_{12} = J_{23} = 0$) and the initial state is a singlet. The data is fit to $\frac{1}{2} \left( 1 + \exp(-(T/T_2^\ast)^c) \right)$, with $c = 1.97 \pm 0.02$ and $T_2^\ast = 3.49 \pm 0.02 \mu$s. The simulations use a coarse-grained time step of $40$ns. (b) Probability of measuring a singlet for the first two spins when $J_{23}/h = 100$MHz, when the initial state is $\ket{S_0} \ket{+}$. The data is fit to $a \exp(-(T/T_2^\ast)^b) \cos(2 \pi J_{23} T/h )+ (1-a)$, with $a = 0.381 \pm 0.001, b = 1.90 \pm 0.04, T_2^\ast = 0.510 \pm 0.004 \mu$s. The simulations use a coarse-grained time step of $5$ns.  In both (a) and (b), $10^3$ independent noise realizations are generated for each time point using Eq.~\eqref{eqt:OUalgorithmMain} with $\Delta t = 10^{-2}$ns. The error bars are $2 \sigma$ confidence intervals calculated using the standard error of the mean. We note that this does not exactly replicate experimental conditions since there may be shot-to-shot correlations.}
   \label{fig:T2starCalibration}
\end{figure}

We can again find analytic expressions under simplifying assumptions.  If we ignore magnetic noise, the probability can be expressed as
\beq
\mathrm{Pr}(t) = \frac{5}{8} + \frac{3}{8} \cos \left( \frac{1}{\hbar} \left( J_{0} t  + J_0  \int_0^t d \tau \xi(\tau) \right) \right) \ .
\eeq
After performing an ensemble average, we obtain
\beq
\mathrm{Pr}(t) = \frac{5}{8} + \frac{3}{8} e^{-J_{0}^2 K(t)} \cos \left( \frac{J_{0} t}{\hbar} \right) \ ,
\eeq
where
\beq \label{eqt:KT2}
K(t) = \frac{1}{2 \hbar^2} \int_0^t dt_1 \int_0^t dt_2 \overline{ \xi(t_1) \xi(t_2)} \ .
\eeq
Because $K(t)$ has the same functional form as in Eq.~\eqref{eqt:KT}, we can similarly identify the $T_2^\ast$ value in the presence of the exchange coupling. We choose parameters that give a $T_2^\ast$ of approximately $0.516 \ \mu$s for $J_0 = 100$MHz.
\begin{enumerate}
\item For the $1/f$ noise model, for the charge noise processes $\left\{ \vec{\xi}_k \right\}_{k=1}^2$, which are unitless, each is given by a sum of 14 independent OU processes with frequencies $\left\{f_k \right\}_{k=1}^{14}$ distributed log-uniformly from $1$ mHz to $10$ GHz, with $p = \left(  2\times 10^{-3} \right)^2$.  This choice of parameters was chosen to be comparable to the model presented in Ref.~\cite{Weinstein2023}, and our simulation results are shown in Fig.~\ref{fig:T2starCalibration}.

\item For the quasi-static noise model, we use $p = (6.099\times 10^{-3})^2$. 

\end{enumerate}

\section{Role of noisy parity-check measurement on parity-flip PSD} \label{app:noisymeasurement}
\begin{figure}[!htbp] %
   \centering
    \includegraphics[width=3.25in]{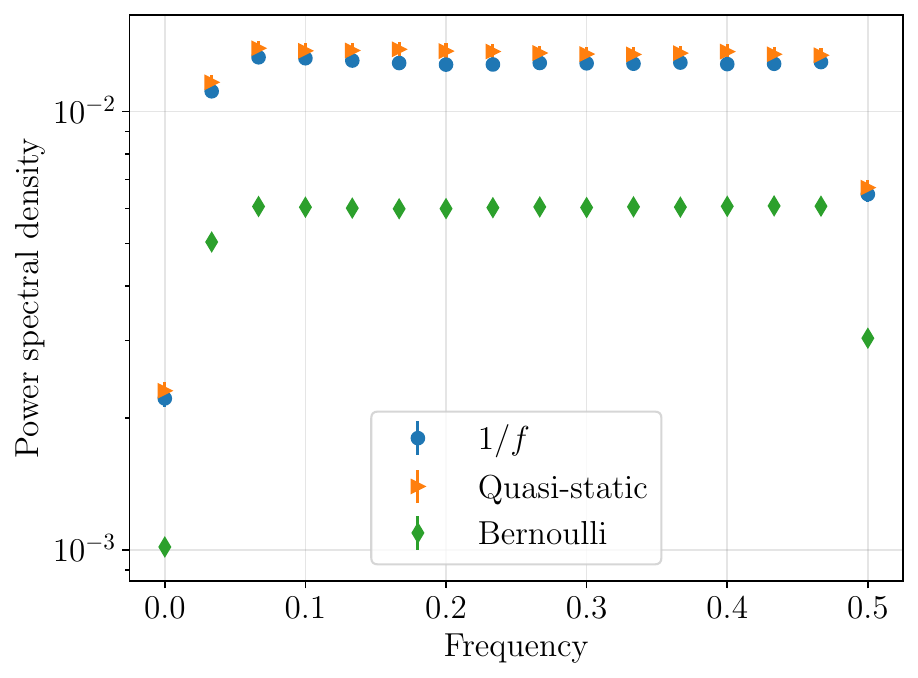} 
   \caption{PSD of the parity-flip time series (Eq.~\eqref{eqt:parityflipTS}) derived from the measurement outcomes for the two simulated noise models and an independent simulation of a Bernoulli process, where the parity measurement is implemented perfectly. The simulated Bernoulli process uses a probability of $q=3 \times 10^{-3}$ and 300 measurements in order to match the other simulations. The PSD is estimated using Welch's method \cite{Welch1967} as implemented in SciPy 1.14.1 \cite{SciPy} using a segment length of 30.  The results are averaged over a total of $4 \times 10^3$ independent simulations, with the error bars  (mostly hidden by the markers) being the $2 \sigma$ confidence interval as inferred by performing a bootstrap over each simulation's estimated PSD.}
   \label{fig:PartiyPSD2}
\end{figure}

We attribute the non-flat PSD of the $1/f$ and quasi-static noise models in Fig.~\ref{fig:PartiyPSD} of the main text to the imperfect parity check measurement that arises from having finite-duration CNOTs and noisy ancilla. To establish this, we perform simulations where the CNOTs are instantaneous and the ancilla qubit is uncorrupted, and we apply the identity operation (see Fig.~\ref{fig:GateImplementation} for our implementation of the identity operation) for the data qubits for 720ns (the same total duration of the two finite-time CNOTs) such that the data qubits can still accumulate errors between parity measurents. We show in Fig.~\ref{fig:PartiyPSD2} the resulting PSD for the parity-flip statistics, where we now observe a flatter PSD in contrast to Fig.~\ref{fig:PartiyPSD}.

\end{document}